\documentclass[showpacs,preprintnumbers,amsmath,amssymb,superscriptaddress,nofootinbib]{revtex4-1}
\usepackage{graphicx}
\usepackage{slashed}

\newcommand{\gsim}{\;\rlap{\lower 3.5 pt \hbox{$\mathchar \sim$}} \raise 1pt
 \hbox {$>$}\;}
\newcommand{\lsim}{\;\rlap{\lower 3.5 pt \hbox{$\mathchar \sim$}} \raise 1pt
 \hbox {$<$}\;}

\usepackage{color}

\allowdisplaybreaks

\begin{document}

\title{Charm-quark mass effects in NRQCD matching coefficients and the
  leptonic decay of the {\boldmath $\Upsilon(1S)$} meson}

\author{Manuel Egner}
\affiliation{Institut f{\"u}r Theoretische Teilchenphysik, Karlsruhe
  Institute of Technology (KIT), 76128 Karlsruhe, Germany}

\author{Matteo Fael}
\affiliation{Institut f{\"u}r Theoretische Teilchenphysik, Karlsruhe
  Institute of Technology (KIT), 76128 Karlsruhe, Germany}

\author{Jan Piclum}
\affiliation{Center for Particle Physics Siegen, Theoretische Physik 1,
  Universit\"at Siegen, 57068 Siegen, Germany}

\author{Kay Sch\"onwald}
\affiliation{Institut f{\"u}r Theoretische Teilchenphysik, Karlsruhe
  Institute of Technology (KIT), 76128 Karlsruhe, Germany}

\author{Matthias Steinhauser}
\affiliation{Institut f{\"u}r Theoretische Teilchenphysik, Karlsruhe
  Institute of Technology (KIT), 76128 Karlsruhe, Germany}


\begin{abstract}
  We compute two-loop corrections to the vector current matching coefficient
  involving two heavy quark masses. The result is applied to the computation of
  the $\Upsilon(1S)$ decay width into an electron or muon pair.  We complement
  the next-to-next-to-next-to-leading order corrections of
  Ref.~\cite{Beneke:2014qea} by charm quark mass effects up to second order in
  perturbation theory.  Furthermore, we apply the formalism to
  $\Gamma(J/\Psi\to \ell^+\ell^-)$ and compare to the experimental data.
\end{abstract}

\preprint{P3H-21-034, SI-HEP-2021-15, TTP21-012}
 
\maketitle



\section{Introduction}

Bottomonia, the bound states of a bottom and an
antibottom quark, are excellent systems to investigate the dynamics of bound
states in QCD. On the experimental side, there exist precise measurements of
their properties. And on the theoretical side, the large mass of the bottom
quark means that perturbation theory can be applied. This is in particular the
case for the $\Upsilon(1S)$ meson. Nevertheless, its description is
complicated by the fact that aside from the bottom-quark mass $m_b$ (the hard
scale), there are two more relevant scales: the typical momentum and energy of
the quarks, which are of order $m_bv$ (the soft scale) and $m_bv^2$ (the
ultrasoft scale), respectively. The $\Upsilon(1S)$ is a non-relativistic bound
state, where the relative velocity $v$ of the quark and antiquark is small,
which means that these scales are well separated. It is then convenient to use
an effective theory for the description of this multiscale problem. Starting
from QCD, we first integrate out the hard modes to arrive at non-relativistic
QCD (NRQCD). In a second step we integrate out the soft modes and potential
gluons with ultrasoft energies and soft momenta to arrive at potential NRQCD
(PNRQCD). At each step one has to determine the Wilson or matching
coefficients of the corresponding effective theory, which are the couplings of
the effective operators. For comprehensive reviews on this topic we refer
to~\cite{Pineda:2011dg,Beneke:2013jia}

The main focus of this paper is the matching coefficient $c_v$ of the vector current
in NRQCD. Among other observables, it contributes to the decay rate of
an $\Upsilon(1S)$ to a lepton-antilepton pair. In PNRQCD and to
next-to-next-to-next-to-leading order (N$^3$LO) accuracy, the decay
rate is given by the formula~\cite{Beneke:2007gj}
\begin{eqnarray}
  \Gamma(\Upsilon(1S)\to \ell^+\ell^-)
  &=& \frac{4\pi\alpha^2}{9m_b^2}
  \left|\psi_1(0)\right|^2
  c_v \left[c_v - \frac{E_1}{m_b} \left(c_v+\frac{d_v}{3}\right) + 
  \ldots\right] \,,\quad
\label{eq:decayrate}
\end{eqnarray}
where $\alpha$ is the fine structure constant and $m_b$ the
bottom-quark pole mass. $E_1$ and $\psi_1(0)$ are the binding energy
and the wave function at the origin of the $(b\bar{b})$ system. For
convenience we provide the leading order results which are given by
\begin{eqnarray}
  \left|\psi_1^{\rm LO}(0)\right|^2 = \frac{8  m_b^3\alpha_s^3}{27 \pi}
  \,,\quad &&
  E_1^{\rm LO}=-\frac{4m_b\alpha_s^2}{9} \,.
\end{eqnarray}
The matching coefficient $c_v$ of the leading current is known at the
three-loop level~\cite{Czarnecki:1997vz,Beneke:1997jm,Marquard:2014pea} for
the case of one massive quark and $n_l$ massless quarks. $d_v$ is the matching
coefficient of the sub-leading $b\bar b$ current in NRQCD. Since it is
multiplied by $E_1$, it is only required at the one-loop level. This result
can be found in Ref.~\cite{Beneke:2013jia}.  Together with the N$^3$LO results
for the energy levels and the wave function at the
origin~\cite{Beneke:2007gj,Beneke:2007pj,Beneke:2013jia}, this made it
possible to evaluate the decay rate at N$^3$LO in Ref.~\cite{Beneke:2014qea}.

One approximation that was made in Ref.~\cite{Beneke:2014qea} was to
treat the charm quark as massless. The aim of this paper is to go
beyond this approximation and include the corrections due to the
charm-quark mass at next-to-next-to-leading order (NNLO). If we consider the
charm-quark mass $m_c$ to be formally of the order of the hard scale $m_b$, the
charm quark has to be integrated out of QCD, leading to NRQCD with two heavy
quarks with different masses. All NRQCD matching coefficients will then
receive contributions due to $m_c$. However, at NNLO only $c_v$ is
affected. Thus we have to compute the fermionic contribution to the two-loop
corrections to $c_v$ for a second non-zero quark mass. The analytic result for
this contribution completes the two-loop evaluation of $c_v$ and together with
its application to $\Gamma(\Upsilon(1S)\to \ell^+\ell^-)$ constitutes the main
result of our paper. 

Another possibility to include the charm-quark mass effects is to consider
$m_c$ to be soft. In this case the charm quark is integrated out of
NRQCD. Then there is no contribution to $c_v$, but instead to the matching
coefficients of PNRQCD, which are the potentials in the Schr\"odinger equation
describing the $(b\bar{b})$ system. At NNLO, only the Coulomb potential is
affected (see Section~3.5 of Ref.~\cite{Beneke:2014pta}). Thus, the $m_c$
dependence then enters in the wave function and binding energy. We will
compare the results of these two approaches.

The remainder of the paper is organized as follows: In the next section we
describe the calculation of $c_v$ and in Section~\ref{sec::caspmc} the
discussion is extended to external axial-vector, scalar and pseudo-scalar
currents, where in addition to the non-singlet also the singlet contributions
have to be considered.  In Section~\ref{sec::Ups1S} we provide updated
predictions for $\Gamma(\Upsilon(1S)\to \ell^+\ell^-)$ and in
Section~\ref{sec::Jpsi} we consider the decay of the $J/\Psi$ and provide
predictions of $\Gamma(J/\Psi \to \ell^+\ell^-)$ up to N$^3$LO.  Our
conclusions are presented in Section~\ref{sec::concl}.  In the Appendix
analytic results for all matching coefficients up to two loops, which are not
presented in the main part of the paper, are provided.  The supplementary
material
to this paper~\cite{progdata} contains computer-readable expressions
of all matching coefficients and all master integrals, which we compute in
this paper.


\section{\label{sec::cvmc}Two-loop matching coefficient for the vector current
  with two masses}

The matching coefficient for the vector current is defined via
\begin{eqnarray}
  j_v^i &=& c_v \tilde{j}_v^i + {\cal O}\left(\frac{1}{m_q}\right)
            \,,
            \label{eq::jv}
\end{eqnarray}
where $m_q$ is the heavy quark mass and 
$j_v^i$ and $\tilde{j}_v^i$ are currents defined in
the full (QCD) and effective (NRQCD) theory. They are given by
\begin{eqnarray}
  j_v^\mu &=& \bar\psi \gamma^\mu \psi\,,\nonumber\\
  \tilde{j}_v^i &=& \phi \sigma^i \chi\,,
\end{eqnarray}
where $\phi$ and $\chi$ are two-component spinors.  Note that in the heavy
quark limit the $0^{\rm th}$ component of $j_v^\mu$ is of order $1/m_q^2$.

A convenient approach to compute $c_v$ is based on the so-called threshold
expansion~\cite{Beneke:1997zp,Smirnov:2002pj} which is applied to the vertex
corrections of a vector current and a heavy quark-antiquark
pair, $\Gamma_v$. Denoting by $Z_2$ the on-shell quark wave function
renormalization constant one obtains the equation~\cite{Beneke:1997jm}
\begin{eqnarray}
  Z_2 \Gamma_v(q_1,q_2) &=& c_v \frac{\tilde{Z}_2}{\tilde{Z}_v}
  \tilde{\Gamma}_v + {\cal O}\left(\frac{1}{m_q}\right) 
  \label{eq::match}
  \,.
\end{eqnarray}
Note that the vector current in QCD has a vanishing anomalous dimension
whereas $\tilde{Z}_v$ deviates form 1 at order $\alpha_s^2$. 
It gets contributions from the colour factors $C_F^2$ and $C_AC_F$
which are not considered in this paper.
The momenta $q_1$ and $q_2$ in Eq.~(\ref{eq::match}) correspond to
the outgoing momenta of the quark and antiquark which are considered
on-shell. Furthermore, we have $(q_1+q_2)^2 = 4m_q^2$, a consequence of the
threshold expansion.

The quantity $\Gamma_v$ is conveniently obtained with the help of
projectors applied to the vertex function $\Gamma^\mu$. It is straightforward
to show that one gets
\begin{eqnarray}
  \Gamma_v &=& \mbox{Tr}\left[ P^{v}_{\mu} \Gamma^{v,\mu} \right]\,,
  \label{eq::proj_Gam}
\end{eqnarray}
with
\begin{eqnarray}
  P^{v}_{\mu} &=& \frac{1}{8 (d-1) m_q^2} \left( -\frac{\slashed{q}}{2} + m_q
  \right) \gamma_\mu \left( \frac{\slashed{q}}{2} + m_q
  \right)\,,
                  \label{eq::proj}
\end{eqnarray}
where $q = q_1+q_2$.

\begin{figure}[t]
  \begin{tabular}{cccc}
  \includegraphics[width=.2\columnwidth]{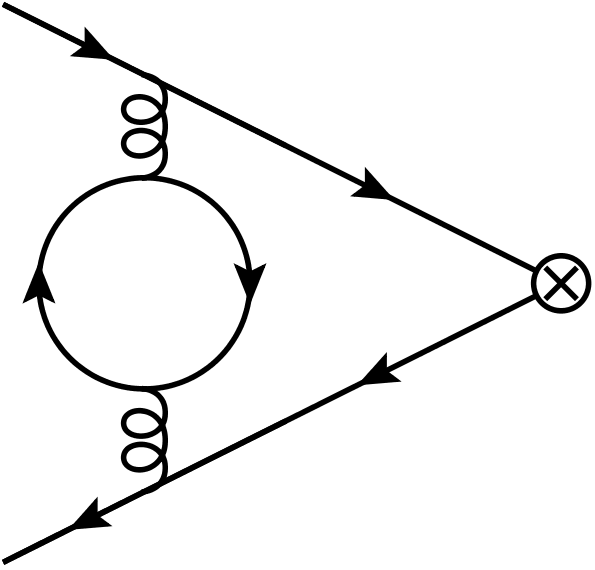} &
  \includegraphics[width=.2\columnwidth]{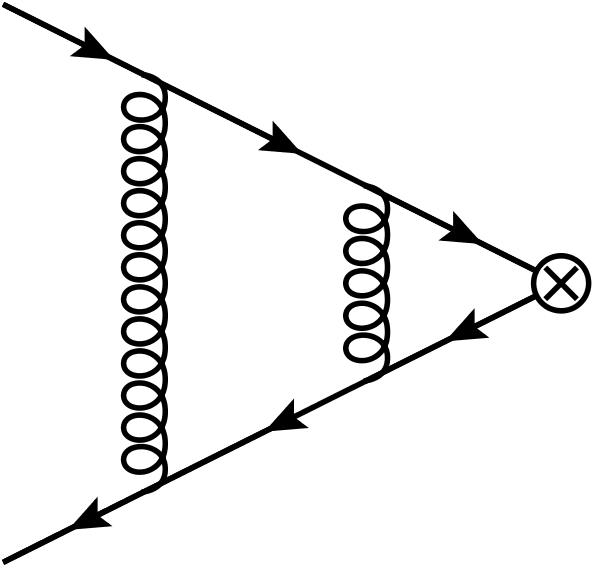} &
  \includegraphics[width=.2\columnwidth]{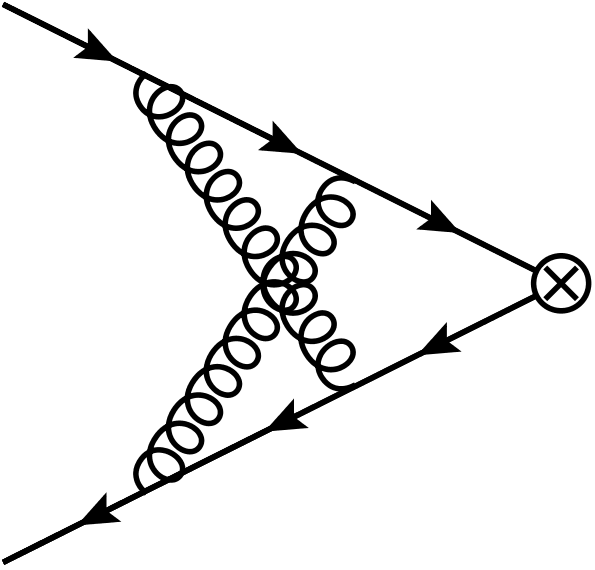} &
  \includegraphics[width=.2\columnwidth]{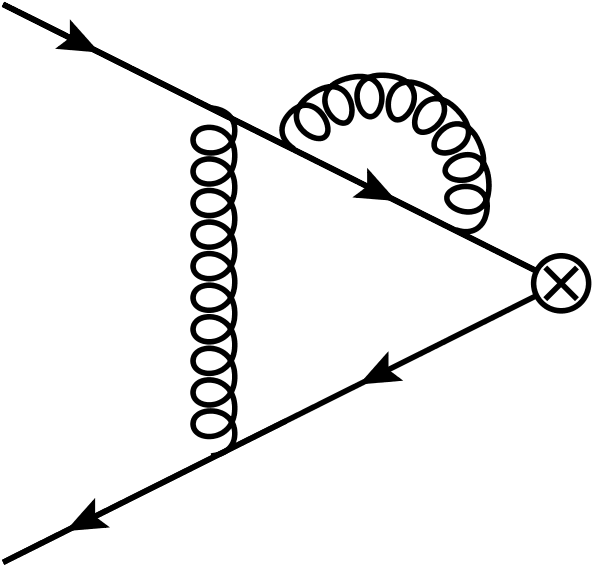} \\
    (a) & (b) & (c) & (d) \\
  \includegraphics[width=.2\columnwidth]{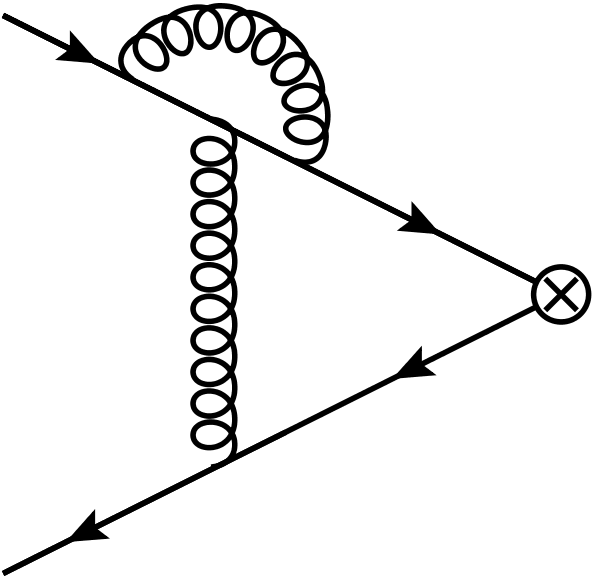} &
  \includegraphics[width=.2\columnwidth]{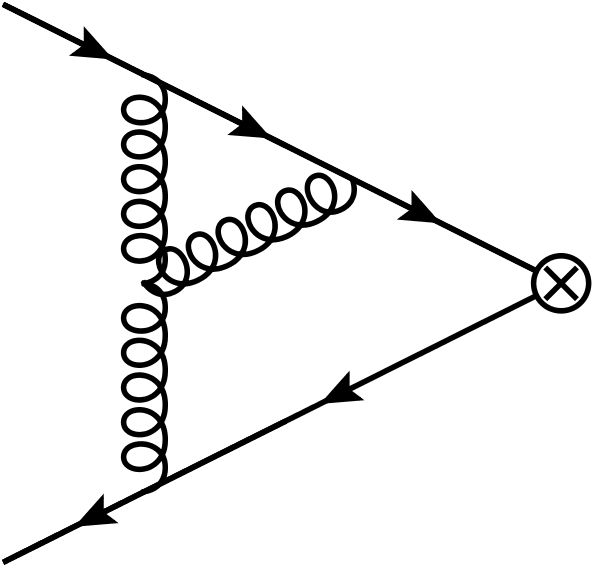} &
  \includegraphics[width=.2\columnwidth]{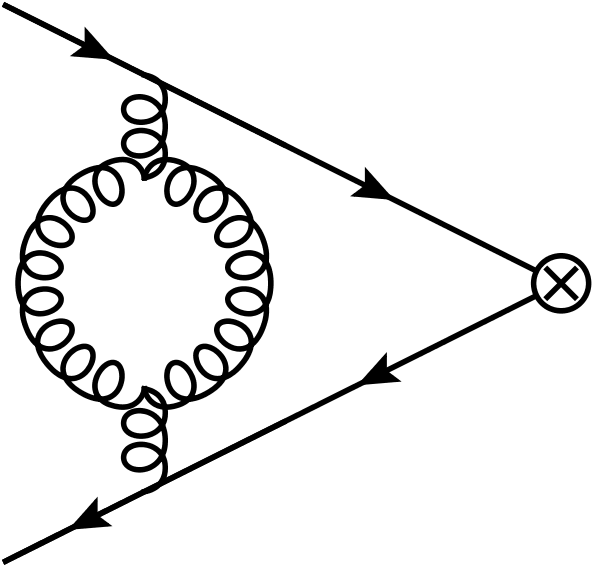} &
  \includegraphics[width=.2\columnwidth]{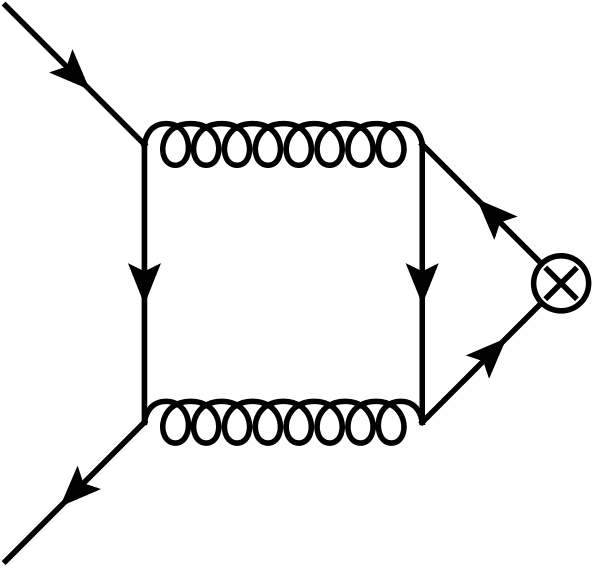}\\
    (e) & (f) & (g) & (h)
  \end{tabular}
  \caption{\label{fig::cv} Sample Feynman diagrams contributions to the
    matching coefficient $c_v$. Straight and curly lines represent quarks and
    gluons, respectively. The cross represents the external current.
    The main focus of this paper is diagram (a) where the quark in the closed 
    loop has mass $m_2$. Note that the singlet diagram shown in (h)
    vanishes for an external vector current. However, for an axial-vector,
    scalar or pseudo-scalar current it is non-zero.}
\end{figure}

\begin{figure}[t]
  \begin{center}
    \includegraphics[width=.8\columnwidth]{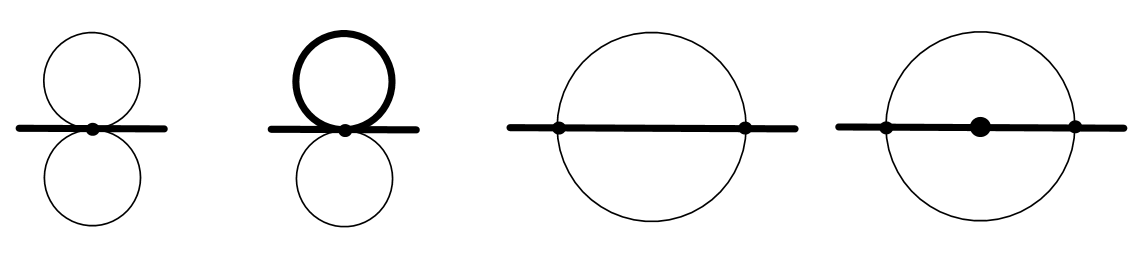}\\
    \mbox{}\hfill $I_1$ \hfill $I_2$ \hfill  $I_3$ \hfill $I_4$ \hfill\mbox{}
    \caption{\label{fig::MIs} Graphical representation of the master integrals
      of Eq.~(\ref{eq::MIs}). Thick and thin solid lines represent
      scalar propagators with mass $m_q$ and $m_2$, respectively. The external
      momentum is always $q/2$ with $(q/2)^2=m_q^2$.
    } 
  \end{center}
\end{figure}

In Fig.~\ref{fig::cv} we show sample diagrams contributing to $c_v$ up to two-loop order.
The main focus of this work is the Feynman diagram in Fig.~\ref{fig::cv}(a)
where the quark in the closed loop has mass $m_2$.
For the computation of this diagram we proceed as follows.
\begin{itemize}

\item We apply the projector in Eq.~(\ref{eq::proj}) to the amplitude of
  the Feynman diagram in Fig.~\ref{fig::cv} and take the traces.
  After decomposing the numerator in terms of denominator factors
  we obtain scalar integrals of the form\footnote{In the denominators we omit $i\varepsilon$
    which could easily be reconstructed by shifting the squared momenta
    according to $p^2\to p^2+i\varepsilon$.}
  \begin{eqnarray}
    I(n_1,\ldots,n_6) = 
    \int 
    \frac{\mathrm{d}^d k}{(2\pi)^d}
    \frac{\mathrm{d}^d l}{(2\pi)^d}
    \frac{ \left(q \cdot l
    \right)^{-n_6}}{\left( -k^2 \right)^{n_1} \left(m_q^2 -
    \left(\frac{q}{2}+k\right)^2 \right)^{n_2} \left(m_q^2 -
    \left(-\frac{q}{2}+k\right)^2 \right)^{n_3} \left(m_2^2 -
    \left(k+l\right)^2 \right)^{n_4} \left(m_2^2 - l^2 \right)^{n_5} }\,. 
    \nonumber\\
  \end{eqnarray}

\item In a next step we perform a partial fraction decomposition
  in order to arrive at integral families where the propagator factors
  are linearly independent. In our case this is achieved with the help
  of
  \begin{eqnarray}
    \int \frac{\mathrm{d}^d k }{ \left( m_q^2 - \left( \frac{q}{2}+k \right)^2
    \right) \left( m_q^2 - \left( -\frac{q}{2}+k \right)^2 \right) } = \int
    \frac{\mathrm{d}^d k}{ \left(-k^2 \right) \left( m_q^2 -
    \left(-\frac{q}{2}+k\right)^2 \right)}\,. 
  \end{eqnarray}

\item We pass the resulting integrals to {\tt FIRE}~\cite{Smirnov:2019qkx} and
  {\tt LiteRed}~\cite{Lee:2012cn} and perform a reduction to four master
  integrals which are given by
  \begin{eqnarray}
    I_1 &=& 
            \int
            \frac{\mathrm{d}^d k}{(2\pi)^d}
            \frac{\mathrm{d}^d l}{(2\pi)^d}
            \frac{1}{
            \left(m_2^2 - k^2 \right) \left(m_2^2 - l^2 \right)
            }\,, \nonumber \\ 
    I_2 &=& 
            \int
            \frac{\mathrm{d}^d k}{(2\pi)^d}
            \frac{\mathrm{d}^d l}{(2\pi)^d}
            \frac{ 1 }{
            \left(m_q^2 - k^2 \right) \left(m_2^2 - l^2
            \right) }\,, \nonumber \\ 
    I_3 &=& 
            \int
            \frac{\mathrm{d}^d k}{(2\pi)^d}
            \frac{\mathrm{d}^d l}{(2\pi)^d}
            \frac{ 1 }{
            \left(m_q^2 - \left(-\frac{q}{2}+k\right)^2 \right) \left(m_2^2 -
            \left(k+l\right)^2 \right) \left(m_2^2 - l^2 \right) }\,,\nonumber \\ 
    I_4 &=& 
            \int
            \frac{\mathrm{d}^d k}{(2\pi)^d}
            \frac{\mathrm{d}^d l}{(2\pi)^d}
            \frac{ 1 }{
            \left(m_q^2 - \left(-\frac{q}{2}+k\right)^2 \right)^2 \left(m_2^2 -
            \left(k+l\right)^2 \right) \left(m_2^2 - l^2 \right) }\,. 
    \label{eq::MIs}
  \end{eqnarray}
  Their graphical representation can be found in Fig.~\ref{fig::MIs}.

\item Next, we introduce the variable $x=m_2/m_q$ and 
  establish differential equations for the master integrals
  of the form
  \begin{eqnarray}
    \frac{{\rm d}\vec{I}}{{\rm d}x} &=& M \cdot \vec{I}\,,
    \label{eq::DE}
  \end{eqnarray}
  where the matrix $M$ decomposes into two $1\times1$ and one 
  $2\times2$ blocks.
  The differential equations are brought to $\epsilon$-form using {\tt CANONICA}~\cite{Meyer:2017joq}:
  \begin{eqnarray}
    \frac{{\rm d}\vec{J}}{{\rm d}x} &=& \epsilon \tilde{M} \cdot \vec{J}\,,
  \end{eqnarray}
  where the matrix $\tilde{M}$ does not depend on $\epsilon$ and
  $\vec{I}=T\cdot\vec{J}$. This allows us to compute $\vec{J}$
  order-by-order in $\epsilon$ and express the result in terms of
  iterated integrals which in our case can be expressed in terms of harmonic
  polylogarithms~\cite{Remiddi:1999ew}.

\item In order to fix the boundary conditions we consider the limits $x\to0$ and
  $x\to1$. This is necessary since in each individual limit some of the
  integration constants drop out. Alternatively it would be possible to
  solve the differential equation to higher order in $\epsilon$.
  The values of $\vec{I}$ for $x=0$ and $x=1$ are used to determine the
  integration constants in $\vec{J}$. 

\end{itemize}

We want to remark that we use a general QCD gauge parameter $\xi$
for our computation. The independence of our final result from $\xi$ is a
welcome cross check. Our results for the master integrals agree with
those given in Appendix~B of Ref.~\cite{Grozin:2020jvt}.

For the renormalization of our two-loop contribution we need the two-loop
corrections to the on-shell wave-function renormalization constant.  The analytic results
for the contribution involving $m_2$ can be found in
Refs.~\cite{Broadhurst:1991fy,Bekavac:2007tk,Davydychev:1998si,Grozin:2020jvt,Fael:2020bgs}.
Furthermore the one-loop counterterm for $\alpha_s$ is needed.  At this point
we use Eq.~(\ref{eq::match}) in order to extract $c_v$.

For the fermionic contributions to $c_v$ it is convenient to introduce
$n_f=n_h+n_m+n_l$, where $n_h=1$ and $n_m=1$ label the contributions with
closed massive quark loops with mass $m_q$ and $m_2$, respectively. $n_l$
counts the massless quarks. Using this notation 
we can cast the result for $c_v$ in the form
\begin{eqnarray}
  c_v &=& 1 -2C_F \frac{\alpha_s^{(n_l+n_m)}(\mu)}{\pi} 
          + \left( \frac{\alpha_s^{(n_l+n_m)}(\mu)}{\pi} \right)^2 c_v^{(2)}
          + {\cal O}(\alpha_s^3)
\end{eqnarray}
where $\alpha_s^{(n_l+n_m)}(\mu)$ is the strong coupling constant where the heavy
quark with mass $m_q$ is decoupled from the running of $\alpha_s$.
The $m_2$-independent contributions to $c_v^{(2)}$ can be found in
Refs.~\cite{Czarnecki:1997vz,Beneke:1997jm,Kniehl:2006qw}. The new
contribution proportional to $n_m$ reads
\begin{eqnarray}
  c_v^{(2)}\Big|_{m_2} 
  &=& n_m C_F T_F \Bigg[
\frac{71}{72}
+\frac{35 x^2}{24}
+\pi ^2 \bigg(
        \frac{3}{32 x}-\frac{11 x}{48}-\frac{17 x^3}{32}+\frac{2 x^4}{9}\bigg)
+\frac{1}{24} \big(
        23+19 x^2\big) H_0
+\frac{4}{3} x^4 H_0^2\nonumber\\&&\mbox{}
+\bigg(
        \frac{3}{16 x}-\frac{11 x}{24}-\frac{17 x^3}{16}+\frac{4 x^4}{3}\bigg) H_0 H_1
+\bigg(
        -\frac{3}{16 x}+\frac{11 x}{24}+\frac{17 x^3}{16}-\frac{4 x^4}{3}\bigg) H_{0,1}\nonumber\\&&\mbox{}
+\bigg(
        \frac{3}{16 x}-\frac{11 x}{24}-\frac{17 x^3}{16}-\frac{4 x^4}{3}\bigg) H_{-1,0}
+\frac{2}{3} \log \bigg(
        \frac{\mu ^2}{m_2^2}\bigg)
                    \Bigg]
      \,,
                    \label{eq::cvm2}
\end{eqnarray}
where $H_{i\ldots}=H_{i\ldots}(x)$ and
$H_{i\ldots}(x)$ are harmonic polylogarithms (HPLs)~\cite{Remiddi:1999ew}.
Note that for $x\to1$ we reproduce the
known result for $m_2=m_q$ which is given by
\begin{eqnarray}
  c_v^{(2)}\Big|_{m_2} & \stackrel{x\to1}{\longrightarrow} 
  & C_F T_F
    \left(\frac{22}{9}-\frac{2}{9}\pi^2 + \frac{2}{3}\log \left(
    \frac{\mu^2}{m_q^2} \right)
    \right)
    \,.
    \label{eq::cvm2_x1}
\end{eqnarray}
However, in the limit $x\to0$ we do not obtain the massless fermion
contribution but recover the well-known Coulomb singularity, which
is regulated by the mass $m_2$. For small $m_2$ we have
\begin{eqnarray}
  c_v^{(2)}\Big|_{m_2} & \stackrel{x\to0}{\longrightarrow} 
  & C_F T_F
    \left( \frac{3 \pi^2}{32 x} + {\frac{11}{18}} 
    + \frac{2}{3}\log \left( \frac{\mu^2}{m_q^2} \right) + {\cal O}(x) \right)
    \,.
    \label{eq::cvm2_x0}
\end{eqnarray}
In the application we discuss in Section~\ref{sec::Ups1S} we need $c_v$
expressed in term of $\alpha_s^{(nl=3)}$ which means that we have to decouple
the charm quark from the running of $\alpha_s$. As a consequence
$\mu^2$ is effectively replaced by $m_2^2$
in Eqs.~(\ref{eq::cvm2}),~(\ref{eq::cvm2_x1})
and~(\ref{eq::cvm2_x0}) disappear.

Let us finally investigate the numerical effect of the new contribution.
We specify to the bottom-charm system and use
$m_2=m_c=1.65$~GeV and $m_q=m_b=5.1$~GeV
for the pole masses of the charm and bottom quarks. This leads to
\begin{eqnarray}
  c_v^{(2)} =  -44.72 + 0.17 n_h + 0.41 n_l + 1.75 n_m + \log \left(
  \frac{\mu^2}{m_q^2} \right) 
  \left[-20.13 + 0.44 (n_l + n_m)\right],
\end{eqnarray}
where the contributions originating from the closed massless, bottom and charm
quark loops are marked by $n_l=3$, $n_h=1$ and $n_m=1$, respectively. 
One observes that the coefficient of $n_m$ is more than a factor four times
larger than the coefficients of $n_l$ and $n_h$. Thus, the
contribution of the heavy quark with mass $m_2$ is larger than the
contributions of the heavy quark with mass $m_q$ and all three massless
quarks combined.


\section{\label{sec::caspmc}Two-loop two-mass matching coefficients for axial-vector, scalar and
  pseudo-scalar currents.}

In this Section we consider further external currents, which are of
phenomenological relevance, and compute the corresponding matching
coefficients between QCD and NRQCD to two-loop order. Such currents have, in
contrast to the vector case, both non-singlet and singlet contributions. The
latter are characterized by the fact that the external current does not
directly couple to the quarks in the final state but only through the exchange of
two gluons. A sample Feynman diagram is shown in Fig.~\ref{fig::cv}(h). 

We write the two-loop corrections in the form
\begin{eqnarray}
  c_x^{(2)} = c^{(2)}_{x, \rm non-sing} + c^{(2)}_{x,\rm sing} 
  \,,
\end{eqnarray}
where $x\in\{a,s,p\}$ stands for an axial-vector, scalar or pseudo-scalar.
All two-loop corrections which involve only one mass scale have been computed
in Ref.~\cite{Kniehl:2006qw}. In this paper we concentrate on the
diagrams where a second massive quark in a closed loop is present
which concerns both the non-singlet and the singlet contribution.

In analogy to Eq.~(\ref{eq::jv}) we define the additional currents in QCD via
\begin{eqnarray}
  j_a^\mu &=& \bar{\psi} \gamma^\mu\gamma_5 \psi\,, \nonumber \\
  j_s     &=& \bar{\psi} \psi\,, \nonumber \\
  j_p     &=& \bar{\psi} i\gamma_5 \psi\,.
  \label{eq::currents}
\end{eqnarray}
The anomalous dimension of $j_a^\mu$ is zero. For the scalar and pseudo-scalar
current we have for the corresponding renormalization constant
$Z_s = Z_p = Z_m$, where $Z_m$ is the (on-shell) mass renormalization
constant.

In NRQCD the currents read~\cite{Kniehl:2006qw}
\begin{eqnarray}
  \tilde{j}_a^i &=& \frac{1}{2 m} \phi^\dagger
  [\sigma^i,\vec{p}\cdot\vec{\sigma}] \chi \,,\nonumber\\
  \tilde{j}_s &=& -\frac{1}{m} \phi^\dagger \vec{p}\cdot\vec{\sigma}
  \chi \,,\nonumber\\
  \tilde{j}_p &=& -i\phi^\dagger \chi \,,
\end{eqnarray}
where $k=1,2,3$. Furthermore we have $j_a^0 = i\tilde{j}_p$ which constitutes
an alternative way to compute the matching coefficient $c_p$. Note the
presence of the momentum $\vec{p}$, which is the relative momentum of the
external quark and antiquark, in the definition of the axial-vector and
scalar current. Thus, an expansion in $p$ has to be performed in order to
obtain the loop corrections to the corresponding matching coefficients.

The matching equation in~(\ref{eq::match}) also holds for the other currents
after the obvious replacements of $\Gamma_v$, $\tilde{Z}_v$ and
$\tilde{\Gamma}_v$.

For the pseudo-scalar current and the zero component of the axial-vector
the momentum $p$ is zero and the calculation proceeds in close analogy to
the vector case. In fact, we have~\cite{Kniehl:2006qw}.
\begin{eqnarray}
  \Gamma_p &=& \mbox{Tr}\left[ P^{(p)} \Gamma^{(p)} \right]\,,
  \nonumber\\
  \Gamma_{a,0} &=& \mbox{Tr}\left[ P^{(a,0)}_{\mu} \Gamma^{(a),\mu} \right]\,,
  \label{eq::proj0}
\end{eqnarray}
with
\begin{eqnarray}
  P^{(p)} &=& \frac{1}{8 m_q^2} \left( -\frac{\slashed{q}}{2} + m_q
  \right) \gamma_5 \left( \frac{\slashed{q}}{2} + m_q
  \right)\,,\nonumber\\
  P^{(a,0)}_{\mu} &=& -\frac{1}{8 m_q^2} \left( -\frac{\slashed{q}}{2} + m_q
  \right) \gamma_\mu \gamma_5 \left( \frac{\slashed{q}}{2} + m_q
  \right)\,.
  \label{eq::proj1}
\end{eqnarray}

For the axial-vector and scalar cases there are similar equations to
Eq.~(\ref{eq::proj0}). The expansion in $p$ (up to linear order) is conveniently
realized by choosing $q_1=q/2+p$ and $q_2=q/2-p$, which implies $q\cdot p =0$.
Thus the projectors are more complicated and are given by
\begin{eqnarray}
  P_{(a,i),\mu} &=& -\frac{1}{8 m_q^2} \left\{ \frac{1}{d-1}
    \left(-\frac{\slashed{q}}{2} + m_q \right) \gamma_\mu \gamma_5
    \left(-\frac{\slashed{q}}{2} + m_q\right) \right. \nonumber\\
  && \left. - \frac{1}{d-2} \left(-\frac{\slashed{q}}{2} + m\right)
    \frac{m}{p^2} \left( (d-3) p_\mu + \gamma_\mu \slashed{p} \right)
    \gamma_5 \left(\frac{\slashed{q}}{2} + m\right) \right\}
    \,,\nonumber\\
  P_{(s)} &=& \frac{1}{8 m_q^2} \left\{ \left(-\frac{\slashed{q}}{2} + m
    \right) {\bf 1} \left(-\frac{\slashed{q}}{2} + m\right) +
    \left(-\frac{\slashed{q}}{2} + m\right) \frac{m}{p^2} \slashed{p}
    \left(\frac{\slashed{q}}{2} + m\right) \right\}
    \,.
  \label{eq::proj2}
\end{eqnarray}
After the application of the projectors and the expansion in $p$ we can set
$p=0$ and $q^2=4m_q^2$. 

The calculation of the non-singlet contribution is in close analogy to the
vector case, see Section~\ref{sec::cvmc}. In particular, it is possible to use
anticommuting $\gamma_5$. Furthermore, we can map the scalar integrals
contributing to $\Gamma_x$ after the application of the projector to the 
same integral families and we thus end up with the same master integrals.

The singlet contribution is more involved and a few comments are in
order. Let us first mention that a non-zero contribution for the scalar and
pseudo-scalar currents
is only obtained for massive quarks in the closed fermion loop.
Furthermore, for the axial-vector current an effective current
formed by the difference of the upper and lower component of a given
quark doublet should be considered in order to guarantee the
cancellation of anomaly-like contributions. For example, for the top-bottom
doublet we have
\begin{eqnarray}
  j_a^\mu &=& \bar{t} \gamma^\mu \gamma_5 t -
  \bar{b} \gamma^\mu \gamma_5 b\,.
\end{eqnarray}
In practice, this means that we have a quark with mass $m_q$ in the
final state and we consider both a massless quark and a quark with mass $m_2$
in the closed quark loop and take the difference.

In the singlet diagrams we treat $\gamma_5$ according to the prescription of
Ref.~\cite{Larin:1993tq}. In the Feynman diagrams we apply for the
axial-vector and pseudo-scalar couplings the replacements
\begin{eqnarray}
  \gamma^\mu\gamma_5 &\to& 
  \frac{i}{3!}\varepsilon^{\mu\nu\rho\sigma} \gamma_\nu\gamma_\rho\gamma_\sigma
  \,,\nonumber\\
  \gamma_5           &\to& 
  \frac{i}{4!}\varepsilon^{\mu\nu\rho\sigma} 
  \gamma_\mu\gamma_\nu\gamma_\rho\gamma_\sigma
  \label{eq::g5}
  \,.
\end{eqnarray}
We perform the same substitution also in the corresponding projectors.  Afterwards
we strip off the two $\varepsilon$ tensors and interpret the product in $d$
dimensions.  This allows us to perform the calculation in close analogy to the
scalar current.

\begin{figure}[t]
    \begin{minipage}{0.19\textwidth}
        \centering
        \includegraphics[width=\textwidth]{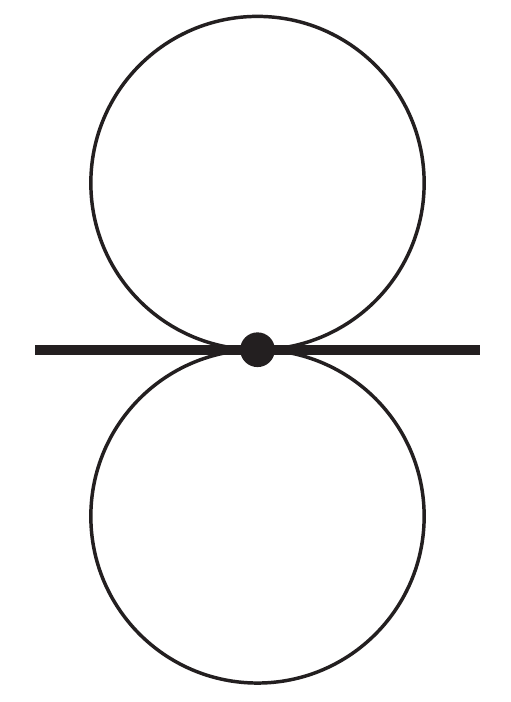}
        $S_1=J_1(0,1,1,0,0)$
    \end{minipage}
    \begin{minipage}{0.19\textwidth}
        \centering
        \includegraphics[width=\textwidth]{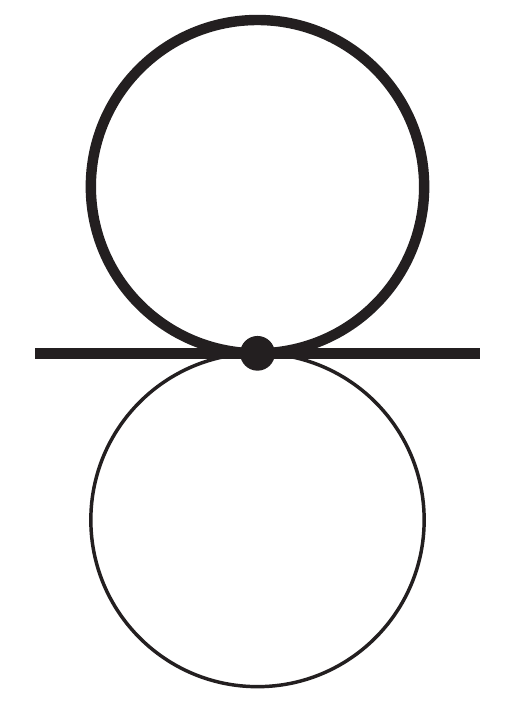}
        $S_2=J_1(0,0,1,1,0)$
    \end{minipage}
    \begin{minipage}{0.3\textwidth}
        \centering
        \includegraphics[width=\textwidth]{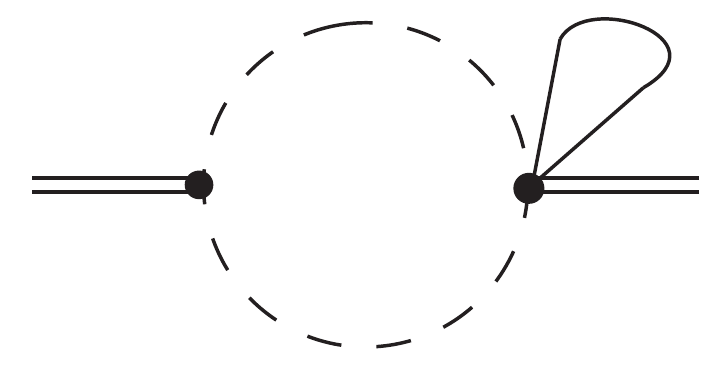}
        $S_3=J_2(0,0,1,1,1)$
    \end{minipage}
    \\
    \begin{minipage}{0.3\textwidth}
        \centering
        \includegraphics[width=\textwidth]{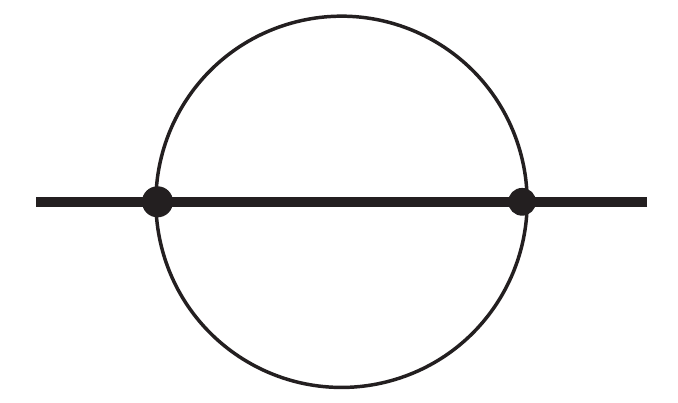}
        $S_4=J_1(0,1,1,1,0)$
    \end{minipage}
    \begin{minipage}{0.3\textwidth}
        \centering
        \includegraphics[width=\textwidth]{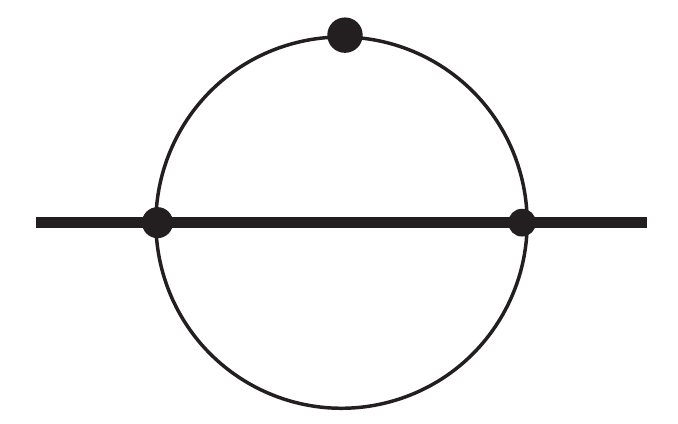}
        $S_5=J_1(0,2,1,1,0)$
    \end{minipage}
    \begin{minipage}{0.3\textwidth}
        \centering
        \includegraphics[width=\textwidth]{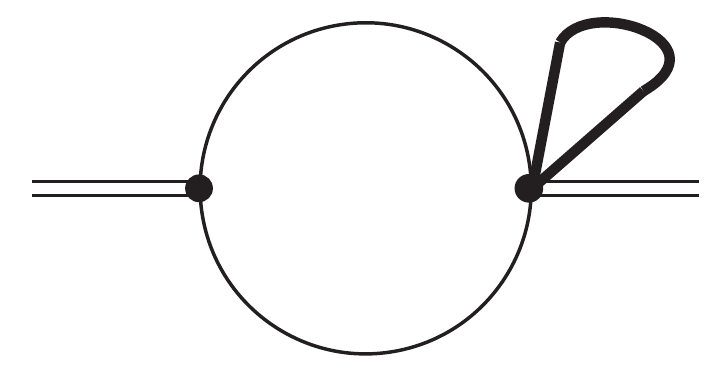}
        $S_6=J_1(1,1,0,1,0)$
    \end{minipage}
    \\
    \begin{minipage}{0.3\textwidth}
        \centering
        \includegraphics[width=\textwidth]{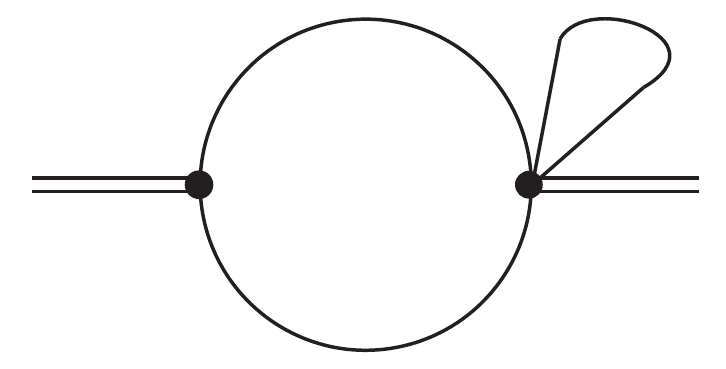}
        $S_7=J_1(1,1,1,0,0)$
    \end{minipage}
    \begin{minipage}{0.3\textwidth}
        \centering
        \includegraphics[width=\textwidth]{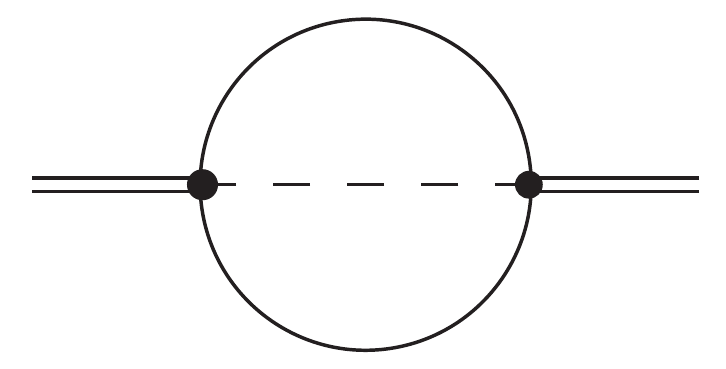}
        $S_8=J_1(0,1,1,0,1)$
    \end{minipage}
    \begin{minipage}{0.3\textwidth}
        \centering
        \includegraphics[width=\textwidth]{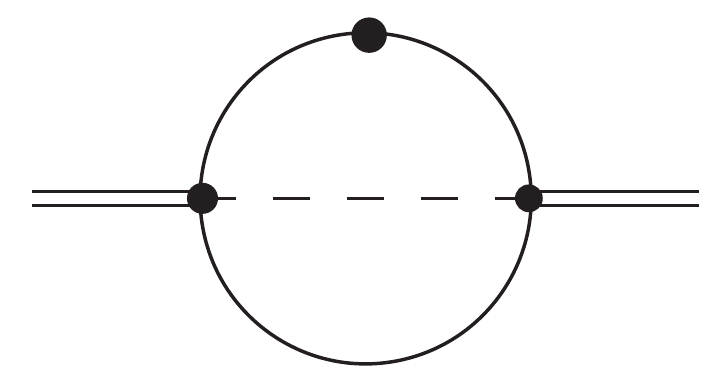}
        $S_9=J_1(0,2,1,0,1)$
    \end{minipage}
    \\
    \begin{minipage}{0.25\textwidth}
        \centering
        \includegraphics[width=\textwidth]{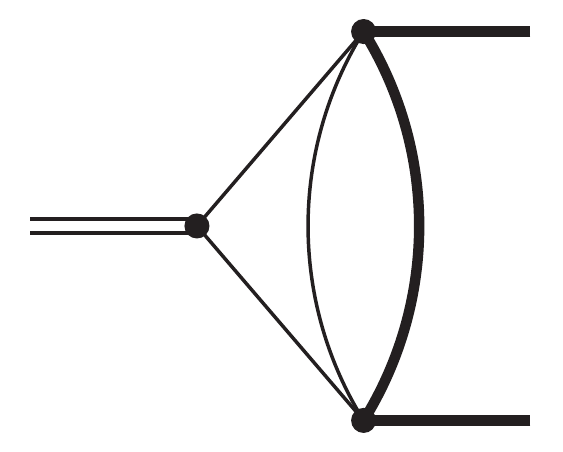}
        $S_{10}=J_1(1,1,1,1,0)$
    \end{minipage}
    \begin{minipage}{0.3\textwidth}
        \centering
        \includegraphics[width=\textwidth]{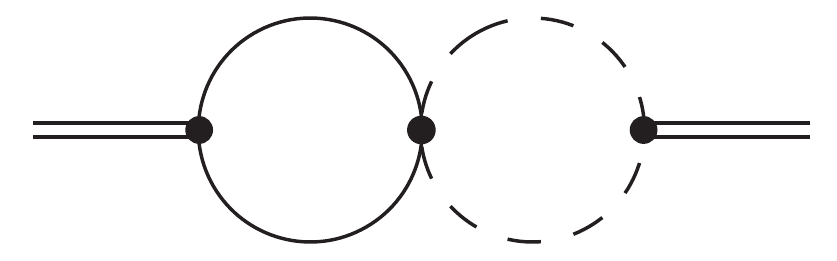}
        $S_{11}=J_2(1,1,0,1,1)$
    \end{minipage}
    \begin{minipage}{0.3\textwidth}
        \centering
        \includegraphics[width=\textwidth]{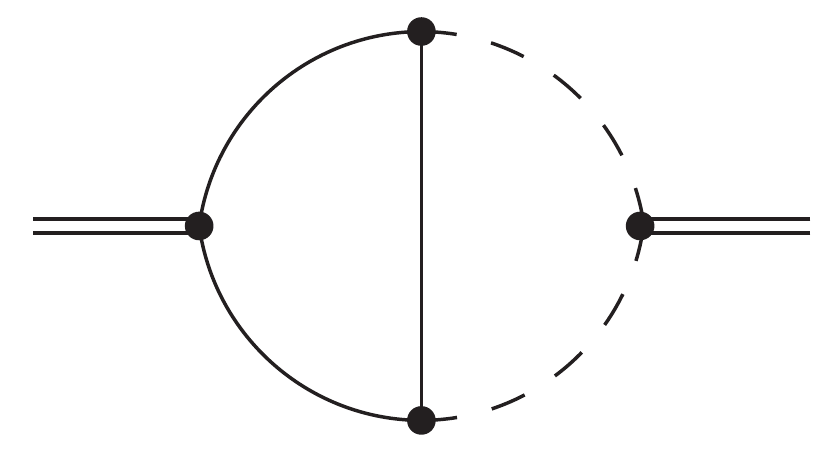}
        $S_{12}=J_2(1,1,1,1,1)$
    \end{minipage}
    \caption{Master integrals needed for the calculation of
    the singlet diagrams. Thick and thin lines represent
    scalar propagators with mass $m_q$ and $m_2$, respectively.
    Dashed lines represent massless scalar propagators.
    External double lines represent the momentum $q$ and
    thick lines the momentum $q/2$ with $q^2=4m_q^2$.}
    \label{fig:mis-singlet}
\end{figure}

The remaining calculation of the singlet diagrams proceeds as outlined in the
previous section.
After applying a partial fraction decomposition, we can map all 
integrals in our amplitude to two integral families which are
given by 
\begin{align}
    J_1(\vec{n}) &= \int \frac{{\rm d}^d k}{(2\pi)^d} \frac{{\rm d}^d l}{(2\pi)^d} 
    \frac{1}{\left(m_2^2-(k+\tfrac{q}{2})^2\right)^{n_1}
    \left(m_2^2-(k-\tfrac{q}{2})^2\right)^{n_2}
    \left(m_2^2-(k-l)^2\right)^{n_3}
    \left(m_q^2 - l^2\right)^{n_4}
    \left(-(l  +  \tfrac{q}{2})^2\right)^{n_5}}
    ~,
    \\
    J_2(\vec{n}) &= \int \frac{{\rm d}^d k}{(2\pi)^d} \frac{{\rm d}^d l}{(2\pi)^d} 
    \frac{1}{\left(m_2^2-(k+\tfrac{q}{2})^2\right)^{n_1}
    \left(m_2^2-(k-\tfrac{q}{2})^2\right)^{n_2}
    \left(m_2^2-(k-l)^2\right)^{n_3}
    \left(-(l+\tfrac{q}{2})^2\right)^{n_4}
    \left(-(l-\tfrac{q}{2})^2\right)^{n_5}}
    ~.
\end{align}
The reduction to master integrals using {\tt FIRE}~\cite{Smirnov:2019qkx} and 
{\tt LiteRed}~\cite{Lee:2012cn} leads to 12 master integrals which are shown
in Fig.~\ref{fig:mis-singlet}.
In a next step we establish differential equations in the variable $t$
defined by $x=2t/(1+t^2)$.
In this new variable the differential equation can be brought into 
$\epsilon$-form with the help of {\tt CANONICA}~\cite{Meyer:2017joq}.
We expand the solution including terms of order $\epsilon$ since 
some of the master integrals have $1/\epsilon$ poles in the prefactor.
The differential equations are integrated with the help of
{\tt HarmonicSums}~\cite{HarmonicSums} in terms of cyclotomic harmonic 
polylogarithms over the alphabet
\begin{align}
    f_0(\tau) &= \frac{1}{\tau}, &
    f_1(\tau) &= \frac{1}{1-\tau}, &
    f_{-1}(\tau) &= \frac{1}{1 + \tau}, &
%
    f_{\{4,1\}}(\tau) &=\frac{\tau}{1+\tau^2}
    .
\end{align}
Alternatively one could factorize the denominators over the 
complex numbers and arrive at Goncharov polylogarithms.
The boundary conditions are obtained from the single-scale master
integrals needed for the two-loop calculation of 
Refs.~\cite{Kniehl:2006qw,Piclum:2007an}.
For the master integrals with dots the naive $m_2=0$ limit is not 
enough and we have to consider the asymptotic expansion around
$m_2=0$ which can be obtained easily from one-dimensional 
Mellin-Barnes representations or a diagrammatic large momentum 
expansion of the corresponding Feynman integrals. 
In a second approach we use the algorithm described 
in~\cite{Ablinger:2018zwz} to solve the differential equations 
in the variable $x$ without going into an $\epsilon$-form first.
For the implementation we additionally make use of 
{\tt Sigma}~\cite{Schneider:2007} and {\tt OreSys}~\cite{ORESYS}.
This approach introduces the square-root valued letter $\sqrt{1-\tau^2}/\tau$.
Both results agree after the above mentioned variable transformation.  We
compute the $\epsilon$ expansion of all master integrals up to the order which
is needed to obtain the ${\cal O}(\epsilon)$ terms of the matching
coefficients.

After inserting the master integrals into the integration-by-parts-reduced 
amplitude we obtain for the two-loop singlet contribution 
to the matching coefficient of the scalar current the following expression
\begin{align}
    c_{s, \rm sing}^{(2)}\Big|_{m_2} &=
    n_m C_F T_F \Biggl[
    \frac{4 t}{3 \big(1+t^2\big)}
    +\pi ^2 \biggl[
         \frac{t \big( 1-28 t^2-35 t^4\big)}{18 \big(1+t^2\big)^3}
        +\frac{4 t^3 H_ 1}{3 \big(1+t^2\big)^3}
        +\frac{2 t \big(3+2 t^2+3 t^4\big) H_{\{4,1\}}}{3 \big(1+t^2\big)^3}
        +\frac{4 t^3 H_{-1}}{\big(1+t^2\big)^3}
    \biggr]
    \nonumber \\ &
    +\log (2) 
    \biggl[
        -\frac{8 t^3 H_ 0}{3 \big(1+t^2\big)^2}
        +\biggl(
            -\frac{4 t}{3 \big(1+t^2\big)}
            +\frac{32 t^3 H_{-1}}{3 \big(1+t^2\big)^3}
        \biggr) H_ 1
        +\frac{16 t^3 H_ 1^2}{3 \big(1+t^2\big)^3}
        +\frac{4 t H_{-1}}{3 \big(1+t^2\big)}
        +\frac{16 t^3 H_ {-1}^2}{3 \big(1+t^2\big)^3}
    \biggr]
    \nonumber \\ &
    +\frac{4 t^3 H_ 0}{\big(1+t^2\big)^2}
    +\frac{4 t^3 \big(3+4 t^2\big) H_ 0^2}{3 \big(1+t^2\big)^3}
    +\biggl(
        \frac{4 t \big(3 t^2-5\big) H_ 0}{3 \big(1+t^2\big)^2}
        +\frac{16 t^3 H_ 0^2}{3 \big(1+t^2\big)^3}
    \biggr) H_ 1
    +\frac{16 t^3 H_ 0 H_ 1^2}{3 \big(1+t^2\big)^3}
    \nonumber \\ &
    +\biggl(
        -\frac{4 t}{1+t^2}
        +\frac{8 t (1-t^2) H_ 0}{3 \big(1+t^2\big)^2}
        -\frac{8 t \big(3+2 t^2+3 t^4\big) H_ 0^2}{3 \big(1+t^2\big)^3}
    \biggr) H_{\{4,1\}}
    +\biggl(
         \frac{4 t \big(5-3 t^2\big)}{3 \big(1+t^2\big)^2}
        -\frac{32 t^3 H_ 0}{\big(1+t^2\big)^3}
        \nonumber \\ &
        -\frac{32 t^3 H_ 1}{3 \big(1+t^2\big)^3}
    \biggr) H_{0,1}
    +\biggl(
        \frac{8 t \big(3 t^2-1\big)}{3 \big(1+t^2\big)^2}
        +\frac{8 t \big(3+2 t^2+3 t^4\big) H_ 0}{\big(1+t^2\big)^3}
    \biggr) H_{0,\{4,1\}}
    +\frac{8 t H_{1,\{4,1\}}}{3 \big(1+t^2\big)}
    \nonumber \\ &
    +\biggl(
         \frac{4 t \big(5-3 t^2\big)}{3 \big(1+t^2\big)^2}
        +\frac{64 t^3 H_ 0}{3 \big(1+t^2\big)^3}
        +\frac{32 t^3 H_ 1}{3 \big(1+t^2\big)^3}
    \biggr) H_{-1,0}
    -\biggl(
         \frac{8 t}{3 \big(1+t^2\big)}
        +\frac{64 t^3 H_ 1}{3 \big(1+t^2\big)^3}
    \biggr) H_{-1,\{4,1\}}
    \nonumber \\ &
    +\frac{160 t^3 H_{0,0,1}}{3 \big(1+t^2\big)^3}
    -\frac{8 t \big(3-2 t+3 t^2\big)\big(3+2 t+3 t^2\big) H_{0,0,\{4,1\}}}{3 \big(1+t^2\big)^3}
    +\frac{32 t^3 H_{0,1,1}}{3 \big(1+t^2\big)^3}
    -\frac{64 t^3 H_{1,1,\{4,1\}}}{3 \big(1+t^2\big)^3}
    \nonumber \\ &
    -\frac{160 t^3 H_{-1,0,0}}{3 \big(1+t^2\big)^3}
    -\frac{32 t^3 H_{-1,0,1}}{3 \big(1+t^2\big)^3}
    +\frac{64 t^3 H_{-1,\{4,1\},1}}{3 \big(1+t^2\big)^3}
    +\frac{32 t^3 H_{-1,-1,0}}{3 \big(1+t^2\big)^3}
    -\frac{64 t^3 H_{-1,-1,\{4,1\}}}{3 \big(1+t^2\big)^3}
    \nonumber \\ &
    -\frac{2 t \log ^2(2)}{3 \big(1+t^2\big)}
    -\frac{t \big(3+4 t^2+3 t^4\big) \zeta (3)}{\big(1+t^2\big)^3}
    \nonumber \\ &
    + i \pi
\Biggl\{
    \frac{2 \pi ^2 t^3}{3 \big(1+t^2\big)^3}
    +\frac{2 t^3}{\big(1+t^2\big)^2}
    +\biggl(
        \frac{4 t^3 \big(4+5 t^2\big)}{3 \big(1+t^2\big)^3}
        +\frac{32 t^3 H_{-1}}{3 \big(1+t^2\big)^3}
    \biggr) H_0
    \nonumber \\ &
    -\biggl(
        \frac{8 t (1-t^2)}{3 \big(1+t^2\big)^2}
        -\frac{16 t^3 H_0}{3 \big(1+t^2\big)^3}
    \biggr) H_1
    +\biggl(
         \frac{4 t (1-t^2)}{3 \big(1+t^2\big)^2}
        -\frac{8 t \big(3+2 t^2+3 t^4\big) H_0}{3 \big(1+t^2\big)^3}
    \biggr) H_{\{4,1\}}
    \nonumber \\ &
    +\frac{8 t (1-t^2) H_{-1}}{3 \big(1+t^2\big)^2}
    -\frac{16 t^3 H_{0,1}}{\big(1+t^2\big)^3}
    +\frac{4 t \big(3+2 t^2+3 t^4\big) H_{0,\{4,1\}}}{\big(1+t^2\big)^3}
    -\frac{16 t^3 H_{-1,0}}{\big(1+t^2\big)^3}
    +\frac{2 t (1-t^2) \log (2)}{3 \big(1+t^2\big)^2}
\Biggr\}
\Biggr]\,,
\label{eq::cs_sing}
\end{align}
with $t=(1-\sqrt{1-x^2})/x$ and $H_{\vec{a}}=H_{\vec{a}}(t)$.
The imaginary part of the matching coefficient is displayed in the last
three lines of Eq.~(\ref{eq::cs_sing}).
For the expansions around $x=0$ we find
\begin{align}
    c_{s,sing}^{(2)}\Big|_{m_2,x\to0} &=
    n_m C_F T_F \biggr[
    x 
    \biggl(
        \frac{2}{3}
        +\frac{\pi ^2}{36}
        -\frac{\log ^2(2)}{3}
        -\frac{3 \zeta (3)}{2}
    \biggr)
    + x^3 
    \biggl(
        -\frac{7}{24}
        -\frac{\pi ^2}{12}
        +\frac{7}{12} \log(x)
        -\frac{3 \log (2)}{4}
        -\frac{1}{3} \log (2) \log(x)
        \nonumber \\ &
        +\frac{\log ^2(2)}{3}
        +\frac{\zeta (3)}{4}
    \biggr)
    + i \pi
    \biggl\{
        \frac{1}{3} x \log (2)
        + x^3 
        \biggl(
            \frac{3}{8}
            +\frac{\pi ^2}{12}
            +\frac{1}{6} \log(x)
            -\frac{\log (2)}{3}
        \biggr)
    \biggr\}
    + \mathcal{O}(x^4)
    \biggr]\,.
\end{align}
As expected, this contribution to the matching coefficient is zero for
vanishing quark mass in the closed
triangle. Note that mass corrections are linear in $m_2$.
On the other hand, for $x\to 1$ $c_{s,sing}^{(2)}\Big|_{m_2}$
approaches a constant. Higher order expansion terms are
conveniently expressed in terms of $y=1-x$ and are given by
\begin{align}
    c_{s,sing}^{(2)}\Big|_{m_2,x \to 1} &=
    n_m C_F T_F
    \biggl[
    \frac{2}{3}
    -\frac{29 \pi ^2}{72}
    -\log (2)
    +\frac{2}{3} \pi ^2 \log (2)
    +y 
    \biggl(
        -\frac{2}{3}
        +\frac{53 \pi ^2}{72}
        +\log (2)-\pi ^2 \log (2)
        -\frac{21 \zeta (3)}{4}
    \biggr)
    \nonumber \\ &
    -\frac{1}{9} \pi ^2 \sqrt{2} y^{3/2}
    +y^2 
    \biggl(
        -\frac{5}{2}
        +\frac{5 \pi ^2}{48}
        +4 \log (2)
        +\frac{1}{2} \pi ^2 \log (2)
        +\frac{63 \zeta (3)}{8}
    \biggr)
    -\frac{7}{30} \pi ^2 \frac{y^{5/2}}{\sqrt{2}}
    \nonumber \\ &
    + i \pi
    \biggl\{
        \frac{1}{2}
        - y 
        \big(
             \frac{1}{2}
            -\frac{\pi ^2}{4}
        \big)
        - y^{3/2} \sqrt{2}
        \biggl(
             \frac{23}{9} 
            -\frac{8}{9}  \log(y)
            -\frac{14}{9}  \log (2)
        \biggr)
        - y^2
        \biggl(
             2
            +\frac{3}{8} \pi ^2
        \biggr) 
        \nonumber \\ &
        + y^{5/2} \sqrt{2}
        \biggl(
            \frac{1591}{300}
            -\frac{6}{5} \log(y)
            -\frac{79}{30} \log (2)
        \biggr)
    \biggr\}
    + \mathcal{O}(y^3)
    \biggr]\,.
\end{align}
The expressions for the pseudo-scalar and axial-vector
currents can be found in Appendix~\ref{App:A}, where
we also show the non-singlet terms.


\section{\label{sec::Ups1S}\boldmath$\Gamma(\Upsilon(1S)\to \ell^+\ell^-)$ and finite charm quark mass}

In Ref.~\cite{Beneke:2014qea} the charm quark has been treated as
massless and the decay rate has been expressed in terms
of $\alpha_s^{(n_l)}(\mu)$ with $n_l=4$. In the following we
discuss the additional ingredients needed for the finite charm quark mass
terms. As mentioned in the Introduction we consider two scenarios:
\begin{itemize}
\item[A.] $m_c$ is hard and the charm quark is integrated out when matching
  QCD to NRQCD. In this approach we express $\Gamma(\Upsilon(1S)\to
  \ell^+\ell^-)$ in terms of $\alpha_s^{(3)}(\mu)$.
  There are finite-$m_c$ effects in the matching coefficient
  $c_v$ starting from two loops. These corrections 
  have been computed in Section~\ref{sec::cvmc}.
  There are no finite-$m_c$ corrections to the binding energy and the wave
  function at the origin.
  
\item[B.] $m_c$ is soft and thus the charm quark is a dynamical scale
  within NRQCD. We express $\Gamma(\Upsilon(1S)\to \ell^+\ell^-)$ in terms of
  $\alpha_s^{(4)}(\mu)$. In this approach
  charm mass effects to bound-state energies and
  wave functions are needed. They are known at NLO~\cite{Eiras:2000rh} and
  NNLO~\cite{Hoang:2000fm,Beneke:2014pta}. We use the expressions
  given in Ref.~\cite{Beneke:2014pta}.

  In case the decay rate shall be expressed in terms of the potential
  subtracted mass the charm quark mass effects are needed to NNLO~\cite{Beneke:2014pta}.

  All necessary expressions for this scenario are available in the program
  {\tt QQbar\_threshold}~\cite{Beneke:2016kkb}.

\end{itemize}
In both scenarios charm mass effects to the
relation between the $\overline{\rm MS}$ (which we use as input) and
on-shell bottom quark mass are taken into account.  They are known to three-loop
order~\cite{Bekavac:2007tk,Fael:2020bgs}.

In scenario~A we assume that $m_c$ is parametrically of the
order of $m_b$. In such a situation both $m_b$ and $m_c$ have
to be decoupled from the running of $\alpha_s$ and $\alpha_s^{(3)}$ is
used as an expansion parameter. 
In fact it has been observed (see, e.g., Ref.~\cite{Ayala:2014yxa}) that,
e.g., the finite-$m_c$ terms to the $\overline{\rm MS}$-on-shell relation of
the bottom quark are quite sizeable and do not converge in case
$\alpha_s^{(4)}$ is used as parameter. On the other hand, charm quark mass
corrections are small and well convergent for $\alpha_s^{(3)}$.

To arrive at the new result for $\Gamma(\Upsilon(1S)\to \ell^+\ell^-)$ we
proceed as follows. Our starting point is the expression derived in
Ref.~\cite{Beneke:2014qea} where $\alpha_s^{(4)}$ has been used as expansion
parameter. For the number of massless quarks we have $n_l=4$.  We restore the
dependence on (massless) charm quarks and write $n_l = n_l^\prime + n_m$ with
$n_l^\prime=3$ and $n_m=1$. In scenario~B we can simply add the
finite-$m_c$ terms from the binding energy and wave function.
This modifies the coefficient of $n_m$ such that in the limit $m_c\to0$
the coefficient of $n_l^\prime$ is recovered.

In scenario~A we interpret the result of Ref.~\cite{Beneke:2014qea}
in the $n_l=3$-flavour PNRQCD with an expansion parameter $\alpha_s^{(3)}$.
Finite-$m_c$ effects enter in Eq.~(\ref{eq:decayrate}) only via
the matching coefficient $c_v$ (cf. Section~\ref{sec::cvmc}) which
also has to be expressed in terms of $\alpha_s^{(3)}$.

We are now in the position to provide numerical results for the decay rate.
For the numerical evaluation we use
$\alpha(2m_b)=1/132.3$~\cite{Jegerlehner:2011mw},
$\alpha_s^{(5)}(M_Z)=0.1179(10)$~\cite{Zyla:2020zbs} and the
renormalization scale $\mu=3.5$~GeV. We use the program {\tt
  RunDec}~\cite{Herren:2017osy} to evolve the coupling with five-loop accuracy
and obtain $\alpha_s^{(4)}(3.5\,\mbox{GeV})=0.2388$
and $\alpha_s^{(3)}(3.5\,\mbox{GeV})=0.2297$, respectively. Furthermore, we compute
the pole mass $m_b=5.059$~GeV in the four-loop approximation from the
$\overline{\mathrm{MS}}$ value $\overline{m}_b(\overline{m}_b)=4.163(16)$~GeV given in
Ref.~\cite{Chetyrkin:2009fv}. In our expressions we renormalize the charm
quark in the $\overline{\rm MS}$ scheme at the renormalization scale
$\mu_c=3$~GeV and use $\overline{m}_c(3~\mbox{GeV})=0.993$~GeV~\cite{Chetyrkin:2009fv}.
Our results in the two scenarios read
\begin{eqnarray}
  \Gamma(\Upsilon(1S)\to \ell^+\ell^-)|_{\mathrm{pole,A}}
  &=& \frac{2^5\alpha^2\alpha_s^3m_b}{3^5}
  \left[1+ 0.374 + (0.916 + 0.020_{c_v}) 
      -0.032 \right] \nonumber \\
  &=& 1.041 + 0.009_{c_v} \nonumber \\
  &=& [1.051\pm 0.047(\alpha_s){}^{+0.007}_{-0.217}(\mu)]~\mbox{keV} \,.
      \nonumber\\
  \Gamma(\Upsilon(1S)\to \ell^+\ell^-)|_{\mathrm{pole,B}}
  &=& \frac{2^5\alpha^2\alpha_s^3m_b}{3^5}
  \left[1+ (0.259 + 0.037_{m_c}) + (0.869 + 0.039_{m_c}) - 0.178 \right] \nonumber\\
  &=&  1.011 + 0.039_{m_c} \nonumber\\
  &=& [1.050\pm 0.045(\alpha_s){}^{+0.024}_{-0.155}(\mu)]~\mbox{keV} \,,
\label{eq:gampole_as4} 
\end{eqnarray}
where the four terms in the first lines of the two equations refer to the LO,
NLO, NNLO and N$^3$LO results. At NNLO and in scenario B also at NLO we
display the contributions from a finite charm quark mass separately. We remark
that the finite-$m_c$ terms of $c_v^{(2)}$, which are computed in
Section~\ref{sec::cvmc}, amount to about $2\%$ of the NNLO coefficient and
they are
of the same order of magnitude as the N$^3$LO contribution.  In scenario~B the
$m_c$ effects at NLO and NNLO are of the same order of magnitude and amount to
about 15\% and 5\% of the corresponding $m_c$-independent coefficient.  The scale
uncertainty in the last line of Eq.~(\ref{eq:gampole_as4}) is computed from
the variation of $\mu$ in the range $\mu\in [3,10]\,\mbox{GeV}$.  We also show
the uncertainty induced by $\delta\alpha_s^{(5)}(M_Z)=0.001$. The variation of
all other parameters leads to significantly smaller uncertainties.

It is interesting to note that both scenario~A and scenario~B lead to the same
final prediction at N$^3$LO although the contributions from the various orders
is different. We oberserve a notable discrepancy to the experimental result
which is given by
$\Gamma(\Upsilon(1S)\to e^+e^-)|_{\mathrm{exp}} = 1.340(18)$~keV.  We also
want to mention a recent lattice evaluation~\cite{Hatton:2021dvg} where the
value $\Gamma(\Upsilon(1S)\to e^+e^-) = 1.292(37)(3)$~keV has been reported.

\begin{figure}[t]
  \includegraphics[width=.8\columnwidth]{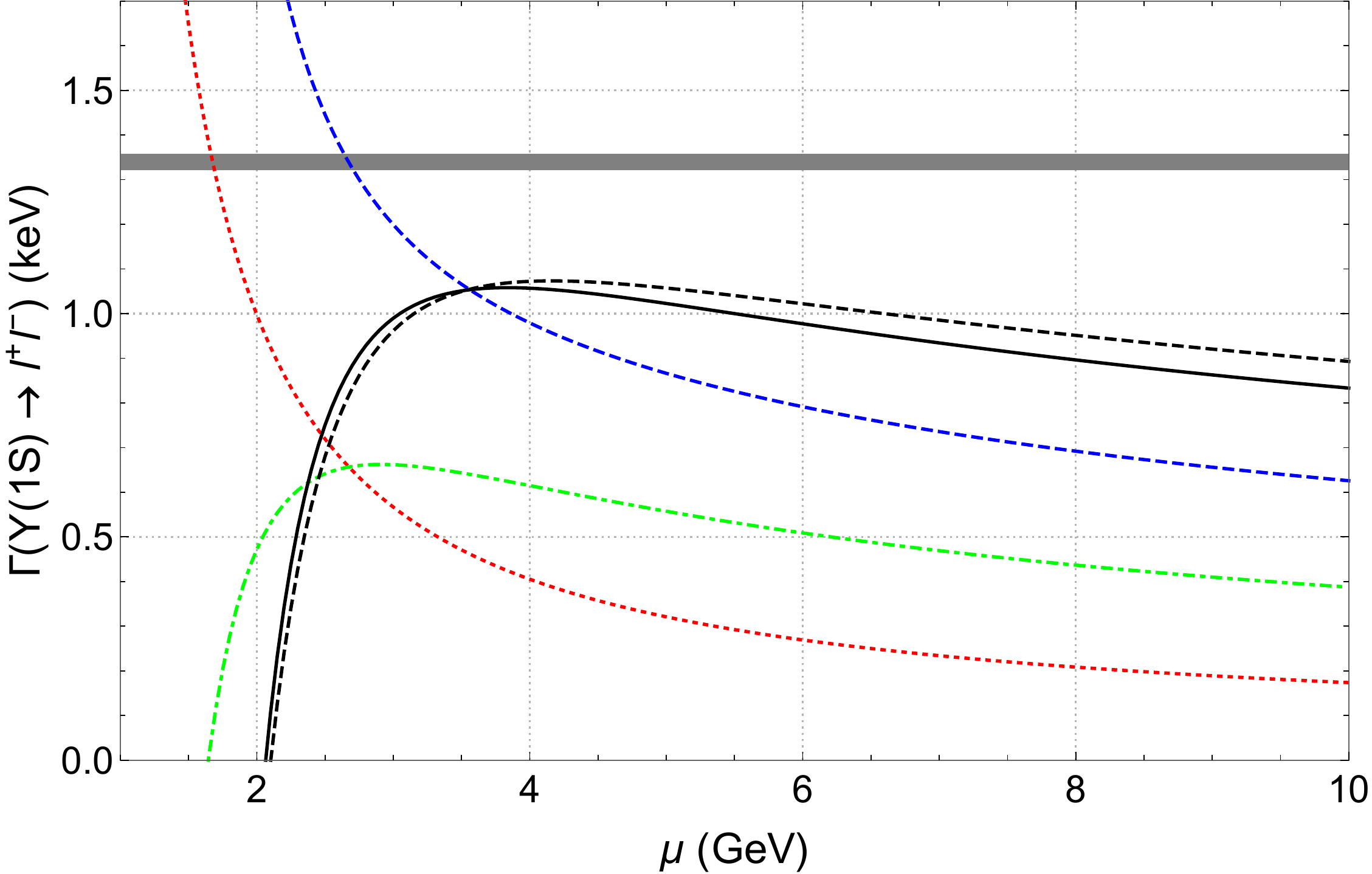}
  \caption{\label{fig::Gamma_as3} The decay rate obtained from scenario~A in
    the OS scheme as a function of the renormalization scale $\mu$. Dotted
    (red), dash-dotted (green), short-dashed (blue) and solid (black) lines
    correspond to LO, NLO, NNLO and N$^3$LO predictions.  At N$^3$LO we also
    show the result from scenario~B as black dashed curve.  The horizontal bar
    denotes the experimental value for $\Gamma(\Upsilon(1S)\to e^+e^-)$. }
\end{figure}

In Fig.~\ref{fig::Gamma_as3} we show the dependence of
$\Gamma(\Upsilon(1S)\to e^+e^-)|_{\mathrm{pole,A}}$ on $\mu$
successively including higher order corrections. The solid black line
corresponds to the N$^3$LO prediction.
We observe that the inclusion of higher order corrections clearly
stabilizes the perturbative predictions for $\mu\gsim 3$~GeV. Furthermore, it is
interesting to note that the third-order corrections vanishes close to the 
value of $\mu$ where the N$^3$LO curve has a maximum.
The dashed black curve corresponds to the N$^3$LO prediction of scenario~B.
The overall shape is very similar to the corresponding curve of scenario~A.
However, it is remarkable that the two N$^3$LO lines cross the
NNLO curve for the same value of $\mu$.

\begin{figure}[t]
  \includegraphics[width=.8\columnwidth]{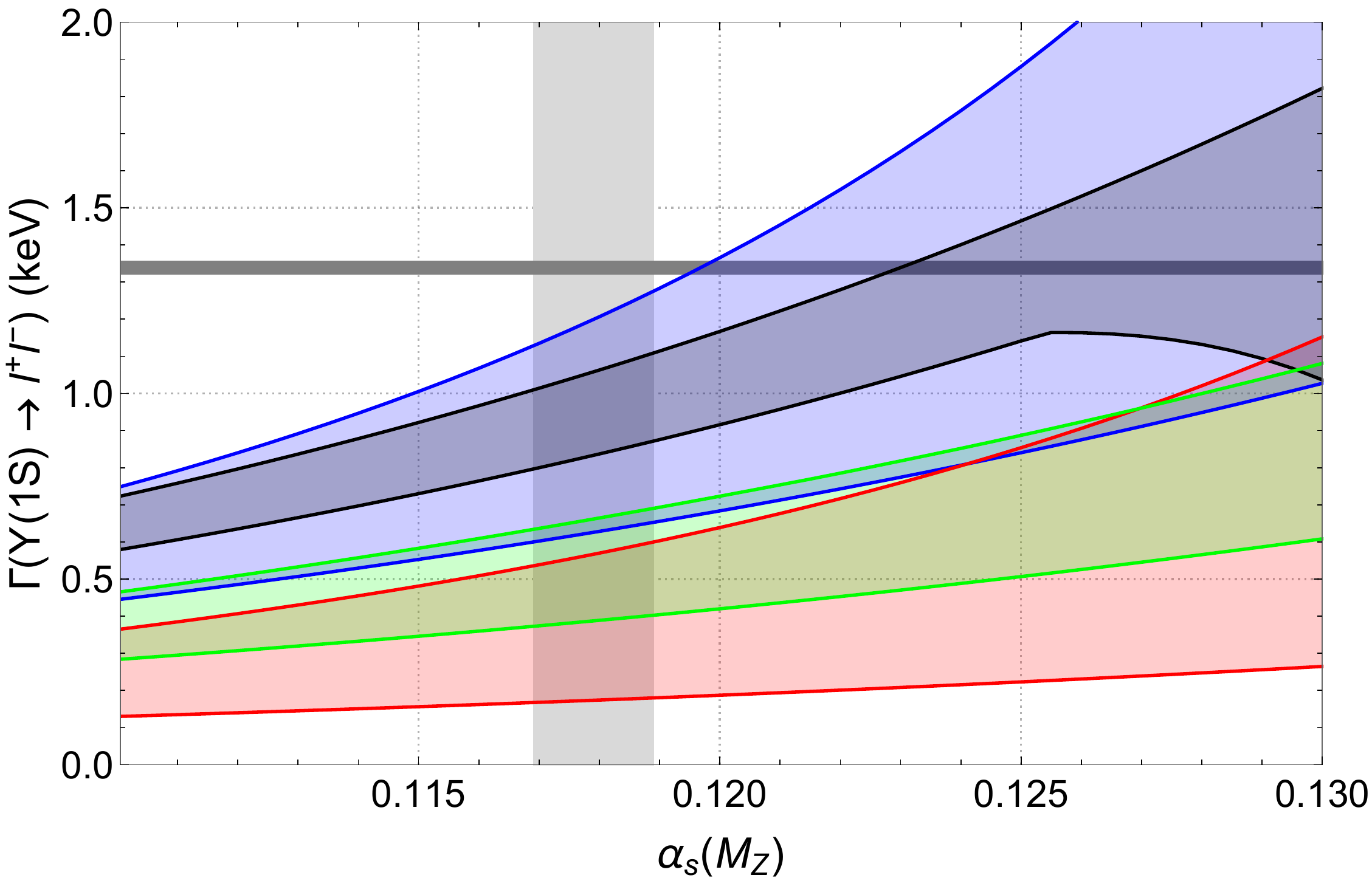}
  \caption{\label{fig::GamAsOS_as3} The decay rate
    obtained from scenario~A in the OS scheme as a
    function of $\alpha_s(M_Z)$ at LO (red, bottom), NLO (green, middle), NNLO
    (blue, top), and N$^3$LO (black, inner top band).  The bands denote the
    variation of $\mu$ between 3~GeV and 10~GeV. The horizontal bar denotes
    the experimental value, while the vertical bar denotes the world average
    of the strong coupling constant, $\alpha_s(M_Z)=0.1179(10)$.}
\end{figure}

In Fig.~\ref{fig::GamAsOS_as3} we show
$\Gamma(\Upsilon(1S)\to e^+e^-)|_{\mathrm{pole,A}}$ as a function of
$\alpha_s^{(5)}(M_Z)$. One observes that the third-order band is
embedded by the NNLO band which can be interpreted as
good convergence of the perturbative corrections.
Note that we do not recompute the bottom pole mass when
varying $\alpha_s$.

It is well-known that the pole mass suffers from so-called renormalon
ambiguities. They are avoided by choosing a properly defined so-called
threshold mass. Such masses have the advantages that they have nice
convergence properties (as the $\overline{\rm MS}$ mass) and that they can also be
used for the description of bound-state properties. In the following we want
to consider the potential-subtracted (PS) mass scheme~\cite{Beneke:1998rk} as
an example and discuss the perturbative corrections to
$\Gamma(\Upsilon(1S)\to\ell^+\ell^-)$.

Explicit results for the relation between the pole mass and the PS mass to
$n$-th order can be derived from the $n$-loop expression for the Coulomb
potential (see for example Ref.~\cite{Beneke:2005hg}). For scenario A, we use
this relation for $n=3$ and $n_l=3$, since in this scenario finite charm-quark
mass effects are only included in the relation between the $\overline{\rm MS}$
mass and pole mass. In scenario B, however, we also have to include charm-mass
effects in the relation between the pole mass and PS mass for $n=1$ and
$n=2$. The latter can be found in Appendix~B of Ref.~\cite{Beneke:2014pta}.

The numerical (input) value for the PS mass is conveniently obtained
from $\bar{m}_b(\bar{m}_b)=4.163$~GeV. Using N$^3$LO accuracy
we obtain for the two scenarios $m_b^{\mathrm{PS}}|_{A}=4.520$~GeV
and $m_b^{\mathrm{PS}}|_{B}=4.484$~GeV, respectively,
where the factorization scale $\mu_f$ is set to 2~GeV.
For the decay rate of the $\Upsilon(1S)$ we obtain
\begin{eqnarray}
  \Gamma(\Upsilon(1S)\to \ell^+\ell^-)|_{\mathrm{PS,A}}
  &=& \frac{2^5\alpha^2\alpha_s^3m_b}{3^5}
  \left[1+ 0.485 + (1.001 + 0.017_{c_v}) +
      0.125 \right] \nonumber \\
  &=& 1.076 + 0.007_{c_v} \nonumber \\
  &=& [1.083\pm 0.053(\alpha_s){}^{+0.001}_{-0.270}(\mu)]~\mbox{keV} \,,
      \nonumber\\
  \Gamma(\Upsilon(1S)\to \ell^+\ell^-)|_{\mathrm{PS,B}}
  &=& \frac{2^5\alpha^2\alpha_s^3m_b}{3^5}
  \left[1+ (0.374 + 0.042_{m_c}) + (0.939 + 0.048_{m_c}) - 0.029 \right] \nonumber \\
  &=& 1.050 + 0.041_{m_c}\nonumber \\
  &=& [1.091\pm 0.052(\alpha_s){}^{+0.006}_{-0.218}(\mu)]~\mbox{keV} \,.
\label{eq:gamPS_as4} 
\end{eqnarray}
The final predictions for the decay rate are close to those in the on-shell
scheme (cf. Eqs.~(\ref{eq:gampole_as4})) and agree well within the
uncertainties. However, the transition from the pole to the PS mass
leads to a significant redistribution among the various perturbative orders.
For example, in scenario~A the N$^3$LO term in the PS scheme is about four
times larger as compared to the on-shell scheme but has a different sign.
Similarly, in scenario~B the N$^3$LO coefficient gets reduced by a factor six.

\begin{figure}[t]
  \includegraphics[width=.8\columnwidth]{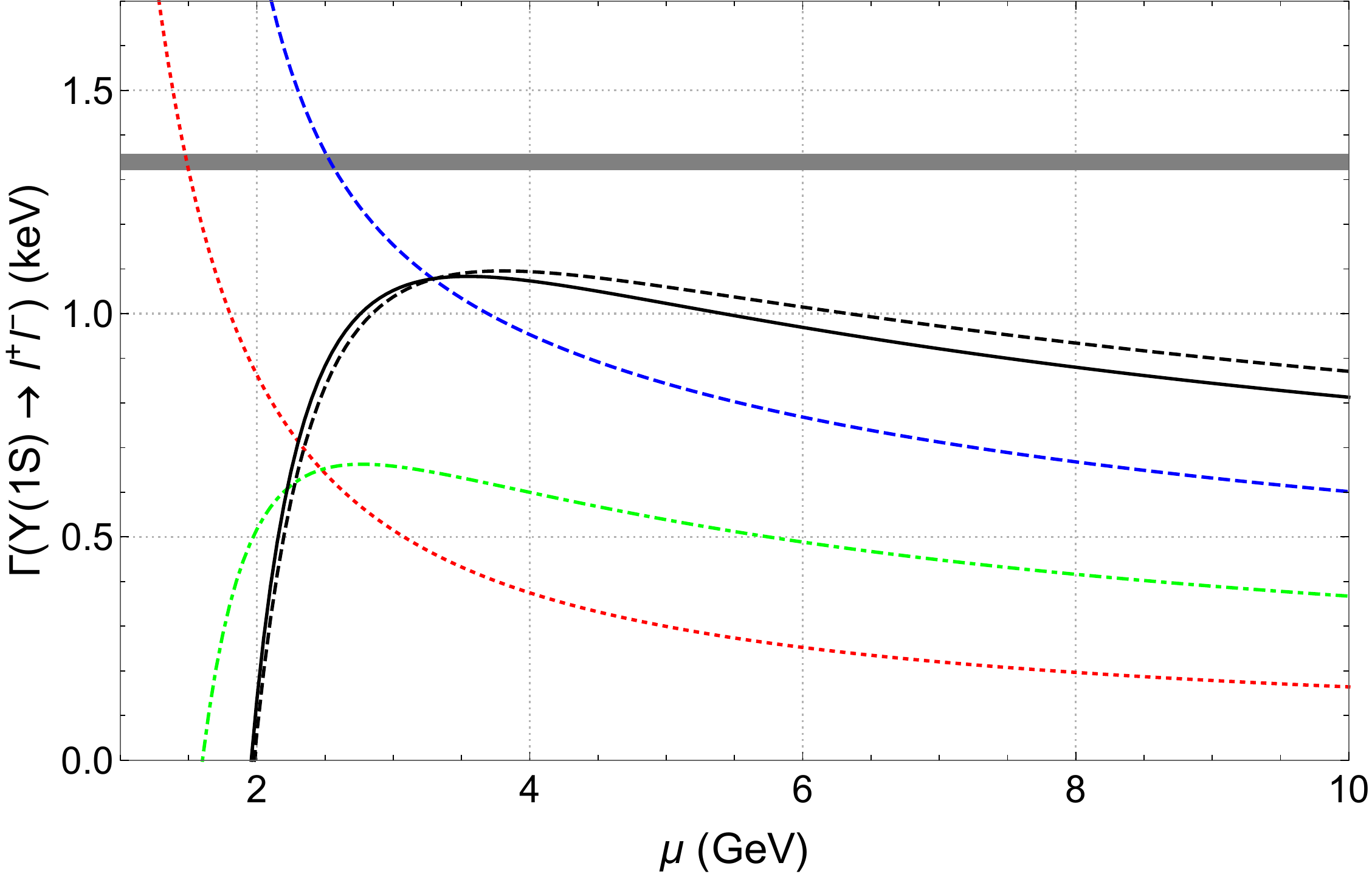}
  \caption{\label{fig::GammaPS_as3} The
    decay rate obtained from scenario~A in the PS scheme as a function of the
    renormalization scale $\mu$. Dotted (red), dash-dotted (green),
    short-dashed (blue) and solid (black) lines correspond to LO, NLO, NNLO
    and N$^3$LO predictions.  At N$^3$LO we also show the result from
    scenario~B as black dashed curve.  The horizontal bar denotes the
    experimental value for $\Gamma(\Upsilon(1S)\to e^+e^-)$. }
\end{figure}

\begin{figure}[t]
  \includegraphics[width=.8\columnwidth]{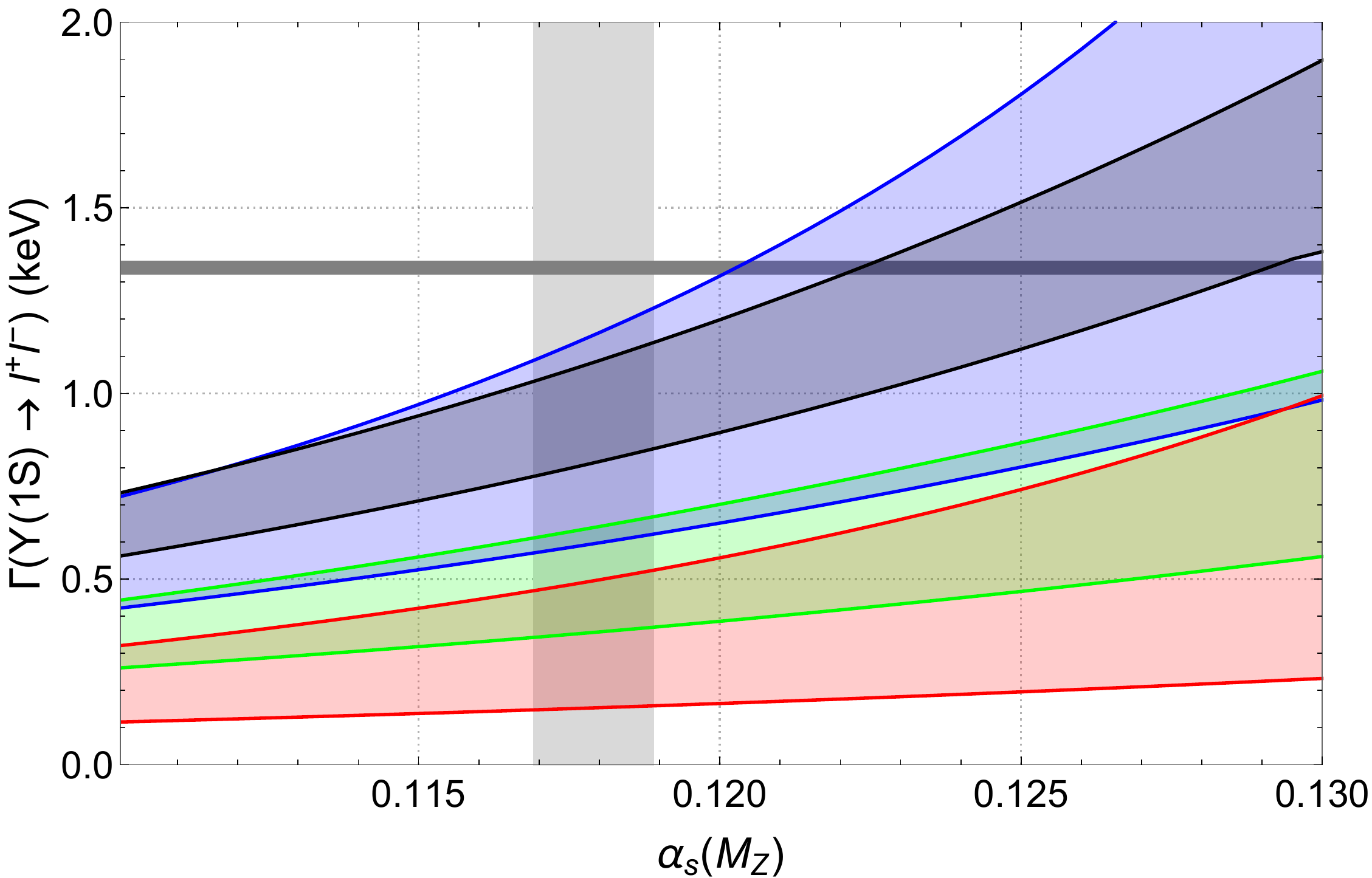}
  \caption{\label{fig::GamAsPS_as3} The decay rate
    obtained from scenario~A in the PS scheme as a
    function of $\alpha_s(M_Z)$ at LO (red, bottom), NLO (green, middle), NNLO
    (blue, top), and N$^3$LO (black, inner top band).  The bands denote the
    variation of $\mu$ between 3~GeV and 10~GeV. The horizontal bar denotes
    the experimental value, while the vertical bar denotes the world average
    of the strong coupling constant, $\alpha_s(M_Z)=0.1179(10)$.}
\end{figure}

For completeness we show in Figs.~\ref{fig::GammaPS_as3}
and~\ref{fig::GamAsPS_as3} the dependence of
$\Gamma(\Upsilon(1S)\to \ell^+\ell^-)$ in the PS scheme on $\mu$ and
$\alpha_s^{(5)}(M_Z)$, respectively.  The behaviour of the various
perturbative orders and the interpretation of the results is very similar to
Figs.~\ref{fig::Gamma_as3} and~\ref{fig::GamAsOS_as3}.

The inclusion of the finite-$m_c$ effects leads to the same conclusions as in
Ref.~\cite{Beneke:2014qea}: The perturbative predictions for
$\Gamma(\Upsilon(1S)\to \ell^+\ell^-)$ are well under control but there is
a discrepancy with respect to the experimental result.  In~\cite{Beneke:2014qea} one can
find an extensive discussion on possible non-perturbative effects. However, no
clear conclusion can be drawn and it remains an open question whether a full
quantitative understanding of the decay rate based on perturbative and
non-perturbative QCD is possible.


\section{\label{sec::Jpsi}\boldmath$\Gamma(J/\Psi\to \ell^+\ell^-)$ at N$^3$LO}

In this section we apply the formalism of Ref.~\cite{Beneke:2014qea} to the
decay of the $J/\Psi$ to massless leptons.  In general, the application to
charm bound states is questionable, since the ultra-soft scale in PNRQCD is
smaller then $\Lambda_{\rm QCD}$. Furthermore, even the hard scale ($m_c$) is
below 2~GeV.  Nevertheless, it is interesting to study the perturbative
behaviour and to compare with the experimental result.

\begin{figure}[t]
  \includegraphics[width=.8\columnwidth]{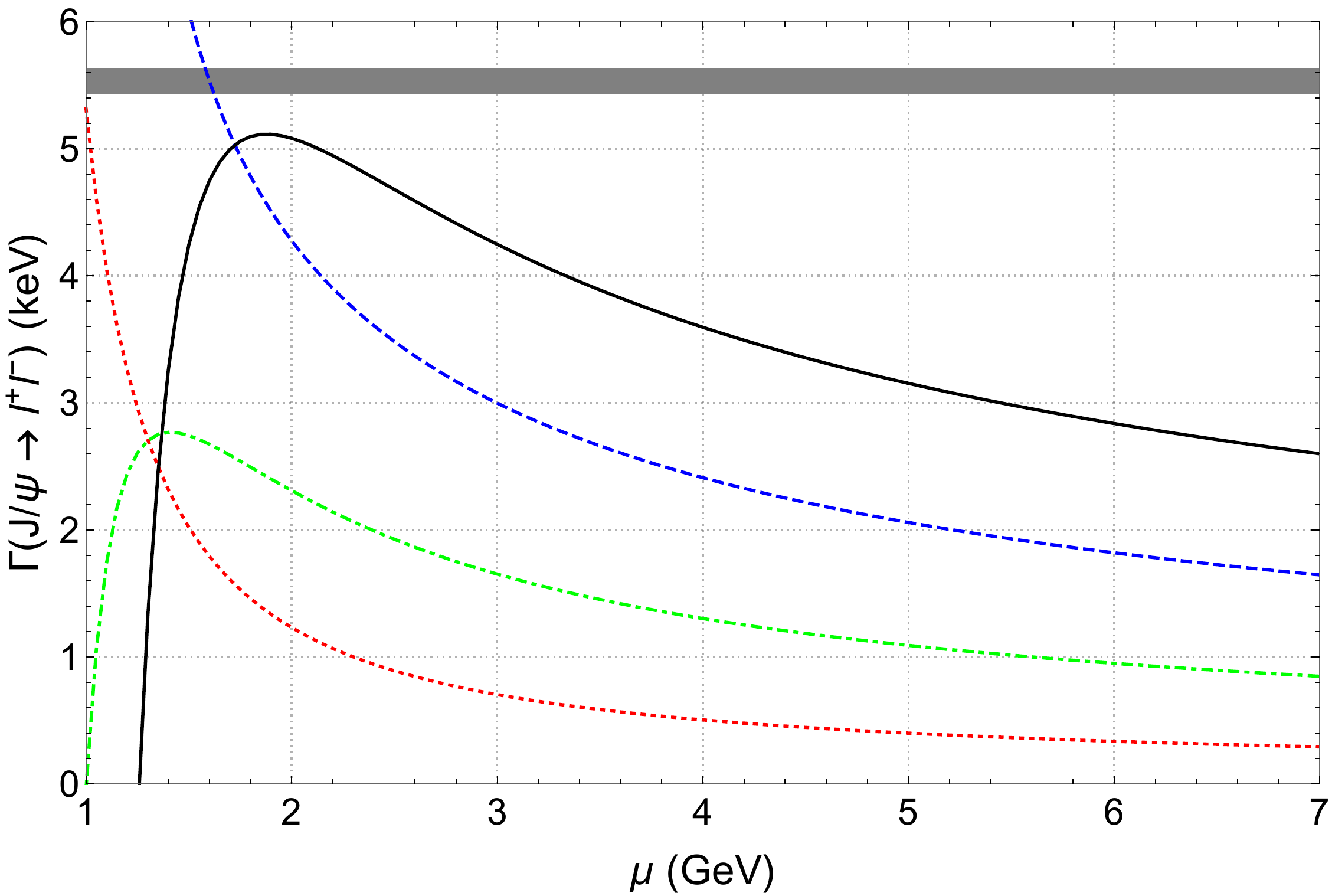}
  \caption{\label{fig::Gamma_jpsi} The decay rate
    $\Gamma(J/\Psi\to \ell^+\ell^-)$ in the OS scheme as a function of the
    renormalization scale $\mu$. Dotted (red), dash-dotted (green),
    short-dashed (blue) and solid (black) lines correspond to LO, NLO, NNLO
    and N$^3$LO prediction.  The horizontal bar denotes the experimental value
    for $\Gamma(J/\Psi\to e^+e^-)$.}
\end{figure}

We work with an on-shell charm quark mass value $m_c=1.65$~GeV
and choose $\mu=2$~GeV for the renormalization scale.  This
leads to $\alpha_s^{(3)}(2\,\mbox{GeV})=0.2943$.  For the decay rate we
have
\begin{eqnarray}
  \Gamma(J/\Psi \to \ell^+\ell^-)|_{\mathrm{pole}}
  &=& \frac{4\,2^5\alpha^2\alpha_s^3m_c}{3^5}
  \left[1 + 0.875 + 1.596 + 0.654 \right] \nonumber \\
  &=& [5.08\pm 0.35(\alpha_s) {}^{+0.03}_{-2.25}(\mu)]~\mbox{keV} \,,
\label{eq:gampole_jpsi} 
\end{eqnarray}
where the scale uncertainty is computed from the variation of $\mu$ in the
range $\mu\in [1.5,6]\,\mbox{GeV}$.  Although the perturbative series does not
converge it is instructive to compare to the experimental result.  This is
done in Fig.~\ref{fig::Gamma_jpsi} where $\Gamma(J/\Psi\to \ell^+\ell^-)$ is
shown as a function of $\mu$.  It is interesting to notice that there is
agreement between the N$^3$LO prediction and the experimental result
$\Gamma(J/\Psi \to \ell^+\ell^-)|_{\mathrm{exp}} = 5.53\pm
0.10$~keV~\cite{Zyla:2020zbs} close to the value of $\mu$ where the N$^3$LO
curve has a maximum and thus the derivative with respect to $\mu$
vanishes. Furthermore, for this value of $\mu$ the third-order corrections are
are quite small, as can be seen from Eq.~(\ref{eq:gampole_jpsi}).  Note that
the N$^3$LO corrections vanish for $\mu=1.724$~GeV. For this
value of the renormalization scale we have
$\Gamma(J/\Psi \to \ell^+\ell^-)|_{\mathrm{pole}} = 5.03$~keV.  From
Fig.~\ref{fig::Gamma_jpsi} we also observe that even the N$^3$LO curve shows a
sizable dependence on $\mu$. Furthermore, one notices that below
$\mu\approx 1.5~\mbox{GeV} \approx m_c$ perturbation theory breaks down.

We want to remark that a similar feature has been observed in
Ref.~\cite{Kniehl:2003ap} where next-to-leading logarithmic (NLL) corrections
to the hyperfine splitting of heavy quark-antiquark bound states have been
considered. The application to the 1S charmonium states shows good agreement
for values of the renormalization scale where the NNL prediction has a
maximum. The perturbative uncertainties are sizeable, as for the $J/\Psi$
decay rate.

A recent lattice computation of the leptonic decay width is given by
$\Gamma(J/\Psi \to e^+ e^-)=5.637\pm0.049$~\cite{Hatton:2020qhk}, in agreement
with the experimental value~\cite{Zyla:2020zbs}.


\section{\label{sec::concl}Conclusions}

In this paper we consider the matching coefficients between QCD and NRQCD of
external vector, axial-vector, scalar and pseudo-scalar currents.  We compute
all two-loop contributions which involve two mass scales, one from the
external quarks and one present in a closed fermion loop.  Whereas for the
vector current only non-singlet contributions have to be considered there are
also singlet contributions for the other three currents.  We present analytic
results including terms of order $\epsilon$, which are of relevance for a
future three-loop calculation.

In Sections~\ref{sec::Ups1S} and~\ref{sec::Jpsi} we apply our results for the
vector current to the leptonic decay rates of the lowest spin-1
heavy-quark-anti-quark mesons, $\Upsilon(1S)$ and $J/\Psi$, and provide update
numerical predictions. We discuss the decay rate
$\Gamma(\Upsilon(1S)\to \ell^+\ell^-)$, including charm quark mass effects,
both in the three- and four-flavour scheme and for the heavy quark masses
defined both in the on-shell and PS scheme.  Although there is a good
convergence of the perturbative corrections, we observe a discrepancy with
respect to the experimental result which to date is not understood.


\section*{Acknowledgements} This research was supported by the Deutsche
Forschungsgemeinschaft (DFG, German Research Foundation) under grant
396021762 --- TRR 257 ``Particle Physics Phenomenology after the Higgs
Discovery''.


\begin{appendix}

\section{\boldmath Analytic results for $c_a$, $c_s$ and $c_p$}
\label{App:A}

In this appendix we present analytic results for the two-mass
matching coefficients for axial-vector, scalar and pseudo-scalar external
currents. The non-singlet results are given by
\begin{eqnarray}
  c_{a, \rm non-sing}^{(2)}\Big|_{m_2} 
  &=& n_m C_F T_F \Bigg[
\frac{37}{72}
+\frac{41 x^2}{24}
+\pi ^2 \bigg(
        \frac{1}{32 x}-\frac{5 x}{48}-\frac{19 x^3}{32}+\frac{2 x^4}{9}\bigg)
+\frac{1}{24} \big(
        13+25 x^2\big) H_0
+\frac{4}{3} x^4 H_0^2 \nonumber\\&&\mbox{}
+\bigg(
        \frac{1}{16 x}-\frac{5 x}{24}-\frac{19 x^3}{16}+\frac{4 x^4}{3}\bigg) H_0 H_1
+\bigg(
        -\frac{1}{16 x}+\frac{5 x}{24}+\frac{19 x^3}{16}-\frac{4 x^4}{3}\bigg) H_{0,1}\nonumber\\&&\mbox{}
+\bigg(
        \frac{1}{16 x}-\frac{5 x}{24}-\frac{19 x^3}{16}-\frac{4 x^4}{3}\bigg) H_{-1,0}
+\frac{1}{3} \log \bigg(
        \frac{\mu ^2}{m_2^2}\bigg)
\Bigg]
      \,,
                    \nonumber\\
c_{s, \rm non-sing}^{(2)}\Big|_{m_2} 
  &=& n_m C_F T_F \Bigg[ 
-\frac{1}{72}
+\frac{27 x^2}{8}
+\pi ^2 \bigg(
        \frac{1}{32 x}+\frac{x}{16}-\frac{35 x^3}{32}+\frac{x^4}{3}\bigg)
+\frac{1}{24} \big(
        5+57 x^2\big) H_0
+2 x^4 H_0^2\nonumber\\&&\mbox{}
+\bigg(
        \frac{1}{16 x}+\frac{x}{8}-\frac{35 x^3}{16}+2 x^4\bigg) H_0 H_1
+\bigg(
        -\frac{1}{16 x}-\frac{x}{8}+\frac{35 x^3}{16}-2 x^4\bigg) H_{0,1}\nonumber\\&&\mbox{}
+\bigg(
        \frac{1}{16 x}+\frac{x}{8}-\frac{35 x^3}{16}-2 x^4\bigg) H_{-1,0}
+\frac{1}{6} \log \bigg(
        \frac{\mu ^2}{m_2^2}\bigg)
   \Bigg]\,, \nonumber\\ 
c_{p, \rm non-sing}^{(2)}\Big|_{m_2} 
  &=& n_m C_F T_F \Bigg[ 
\frac{11}{24}
+\frac{25 x^2}{8}
+\pi ^2 \bigg(
        \frac{3}{32 x}-\frac{x}{16}-\frac{33 x^3}{32}+\frac{x^4}{3}\bigg)
+\frac{1}{8} \big(
        5+17 x^2\big) H_0
+2 x^4 H_0^2\nonumber\\&&\mbox{}
+\bigg(
        \frac{3}{16 x}-\frac{x}{8}-\frac{33 x^3}{16}+2 x^4\bigg) H_0 H_1
+\bigg(
        -\frac{3}{16 x}+\frac{x}{8}+\frac{33 x^3}{16}-2 x^4\bigg) H_{0,1}\nonumber\\&&\mbox{}
+\bigg(
        \frac{3}{16 x}-\frac{x}{8}-\frac{33 x^3}{16}-2 x^4\bigg) H_{-1,0}
+\frac{1}{2} \log \bigg(
        \frac{\mu ^2}{m_2^2}\bigg)
\Bigg]\,,
                    \label{eq::caspm2_nonsing}
\end{eqnarray}
where $H_{\vec{a}}=H_{\vec{a}}(x)$.

Our results for the pseudo-scalar singlet contribution reads
\begin{eqnarray} 
  c_{p, \rm sing}^{(2)}\Big|_{m_2} 
  &=& n_m C_F T_F \Bigg[ 
\pi ^2 \bigg(
        \frac{7 t^3}{3 \big(
                1+t^2\big)^3}
        +\frac{2 t H_{\{4,1\}}}{1+t^2}
\bigg)
-\frac{4 t^3 H_0}{\big(
        1+t^2\big)^2}
+\frac{4 t^3 H_0^2}{\big(
        1+t^2\big)^3}
+\frac{16 t^3 H_0 H_1}{\big(
        1+t^2\big)^3} \nonumber\\&&\mbox{}
+\log (2) \bigg(
        -\frac{2 t}{1+t^2}
        +\frac{16 t^3 H_0}{\big(
                1+t^2\big)^3}
        +\frac{16 t^3 H_1}{\big(
                1+t^2\big)^3}
        -\frac{16 t^3 H_{-1}}{\big(
                1+t^2\big)^3}
\bigg)
+\bigg(
        \frac{4 t}{1+t^2}
        -\frac{8 t H_0^2}{1+t^2}
\bigg) H_{\{4,1\}}\nonumber\\&&\mbox{}
-\frac{16 t^3 H_{0,1}}{\big(
        1+t^2\big)^3}
+\bigg(
        -\frac{32 t^3}{\big(
                1+t^2\big)^3}
        +\frac{24 t H_0}{1+t^2}
\bigg) H_{0,\{4,1\}}
-\frac{32 t^3 H_{1,\{4,1\}}}{\big(
        1+t^2\big)^3}
-\frac{16 t^3 H_{-1,0}}{\big(
        1+t^2\big)^3}\nonumber\\&&\mbox{}
+\frac{32 t^3 H_{-1,\{4,1\}}}{\big(
        1+t^2\big)^3}
-\frac{24 t H_{0,0,\{4,1\}}}{1+t^2}
+\frac{8 t^3 \log ^2(2)}{\big(
        1+t^2\big)^3}
-\frac{3 t \zeta (3)}{1+t^2}\nonumber\\&&\mbox{}
+i \pi \Biggl\{
-\frac{(-1+t) t (1+t)}{\big(
        1+t^2\big)^2}
-\frac{4 t^3 H_0}{\big(
        1+t^2\big)^3}
-\frac{8 t H_0 H_{\{4,1\}}}{1+t^2}
+\frac{12 t H_{0,\{4,1\}}}{1+t^2}
\Biggr\}
\Bigg]\,.
                    \label{eq::caspm2_sing}
\end{eqnarray}
For the expansions around $x=0$ and $x=1$ we find
\begin{eqnarray}
 c_{p, {\rm sing}}^{(2)}\Big|_{m_2,x \to 0} &=& 
        n_m C_F T_F \Bigg[
        x 
        \bigg(
                - \log (2)
                - \frac{3 \zeta (3)}{2} 
        \bigg)
        + x^3 
        \bigg(
                -\frac{1}{8}
                +\frac{5 \pi ^2}{12}
                -\frac{\log (2)}{4}
                -\log ^2(2)
                +\frac{1}{4} (1+8 \log (2)) \log (x)
        \bigg)  
        \nonumber\\&&\mbox{}
        + i \pi 
        \biggl\{ 
                \frac{1}{2}x 
                + x^3 \bigg( \frac{1}{8} +\log(2) -\log(x)\bigg)
        \biggr\} 
        + \mathcal{O}(x^4)
        \Bigg]
        \,, \\
        c_{p,{\rm sing}}^{(2)}\Big|_{m_2,x \to 1} &=&
        n_m C_F T_F \Bigg[
        \frac{5 \pi ^2}{24}
        +\frac{1}{2} \pi ^2 \log (2)
        -\frac{21 \zeta (3)}{8}  
        -\sqrt{y} \frac{\pi ^2}{\sqrt{2}} 
        +y 
        \bigg(
                -1 
                -\frac{\pi ^2}{8}
                +2 \log(2)
                -\frac{1}{2} \pi ^2 \log (2)
                +\frac{21 \zeta (3)}{8}
        \bigg)  
        \nonumber\\&&\mbox{}
        + y^{3/2}\frac{7 \pi ^2}{12\sqrt{2}}
        +y^2 
        \bigg(
                \frac{1}{3}
                +\frac{3 \pi ^2}{8}
                -\frac{\log (2)}{3}
        \bigg)
        +y^{5/2}\frac{71\pi^2}{480\sqrt{2}} 
        \nonumber\\&&\mbox{}
        +i \pi 
        \biggl\{
                \frac{ \pi ^2}{8} +
                \sqrt{y}\bigg(   
                        \sqrt{2}
                        - \sqrt{2} \log (2) 
                \bigg) 
                + y 
                \bigg(
                        -1 
                        -\frac{\pi ^2}{8} 
                \bigg) 
                +y^{3/2}
                \bigg( 
                        -\frac{11}{6\sqrt{2}}  
                        +\frac{7}{6\sqrt{2}}\log (2) 
                \bigg) 
                +\frac{y^2}{6} 
                \nonumber\\&&\mbox{}
                +y^{5/2}
                \bigg( 
                        \frac{433}{240\sqrt{2}} 
                        + \frac{71}{240 \sqrt{2}} \log (2) 
                \bigg)
        \biggr\} + \mathcal{O}(y^3)
        \Bigg]\,,
\end{eqnarray}
with $y=1-x$.

In the case of the singlet axial-vector current we explicitly
specify the flavour of the quark in the final state to bottom quark.
Furthermore, we split the matching coefficient into the
contributions from the strange and charm quarks
$c_{a, \rm sing}^{(2),s+c}\Big|_{m_2}$ and the contribution from the bottom
and top quarks $c_{a, \rm sing}^{(2),b+t}\Big|_{m_2}$.\footnote{Note that the
  contribution from up and down quarks vanishes since we assume that both
  quarks are massless.} 
For vanishing strange quark mass we have
\begin{eqnarray}
        c_{a, \rm sing}^{(2),s+c}\Big|_{m_2} 
        &=& n_m C_F T_F \Bigg[
      \pi ^2 \bigg(
              -
              \frac{t^2 \big(
                      1+25 t^2+19 t^4+9 t^6\big)}{9 \big(
                      1+t^2\big)^4}
              +\frac{2 t^2 H_1}{3 \big(
                      1+t^2\big)^2}
              +\frac{8 t^2 H_{\{4,1\}}}{3 \big(
                      1+t^2\big)^2}
              +\frac{2 t^2 H_{-1}}{\big(
                      1+t^2\big)^2}
      \bigg)\nonumber\\&&\mbox{}
      +\log (2) \bigg(
              \frac{4 t^2}{3 \big(
                      1+t^2\big)^2}
              -\frac{8 t^4 \big(
                      5+2 t^2+t^4\big) H_0}{3 \big(
                      1+t^2\big)^4}
              +\bigg(
                      -\frac{4 \big(
                              1+2 t^2+10 t^4+2 t^6+t^8\big)}{3 \big(
                              1+t^2\big)^4}
                      +\frac{16 t^2 H_{-1}}{3 \big(
                              1+t^2\big)^2}
              \bigg) H_1\nonumber\\&&\mbox{}
              +\frac{8 t^2 H_1^2}{3 \big(
                      1+t^2\big)^2}
              +\frac{4 \big(
                      1+2 t^2+10 t^4+2 t^6+t^8\big) H_{-1}}{3 \big(
                      1+t^2\big)^4}
              +\frac{8 t^2 H_{-1}^2}{3 \big(
                      1+t^2\big)^2}
      \bigg)
      +\frac{4 t^2 \big(
              3+8 t^2+3 t^4\big) H_0}{3 \big(
              1+t^2\big)^3}\nonumber\\&&\mbox{}
      +\frac{4 t^4 \big(
              -1+2 t^2+t^4\big) H_0^2}{3 \big(
              1+t^2\big)^4}
      +\bigg(
              \frac{4 \big(
                      -3-6 t^2-10 t^4+2 t^6+t^8\big) H_0}{3 \big(
                      1+t^2\big)^4}
              +\frac{8 t^2 H_0^2}{3 \big(
                      1+t^2\big)^2}
      \bigg) H_1\nonumber\\&&\mbox{}
      +\frac{8 t^2 H_0 H_1^2}{3 \big(
              1+t^2\big)^2}
      +\bigg(
              -
              \frac{4 \big(
                      -3-6 t^2-10 t^4+2 t^6+t^8\big)}{3 \big(
                      1+t^2\big)^4}
              -\frac{16 t^2 H_0}{\big(
                      1+t^2\big)^2}
              -\frac{16 t^2 H_1}{3 \big(
                      1+t^2\big)^2}
      \bigg) H_{0,1}\nonumber\\&&\mbox{}
      +\bigg(
              -\frac{4 \big(
                      3+8 t^2+3 t^4\big)}{3 \big(
                      1+t^2\big)^2}
              -\frac{32 t^2 H_0^2}{3 \big(
                      1+t^2\big)^2}
      \bigg) H_{\{4,1\}}
      +\bigg(
              \frac{16 t^4 \big(
                      5+2 t^2+t^4\big)}{3 \big(
                      1+t^2\big)^4}
              +\frac{32 t^2 H_0}{\big(
                      1+t^2\big)^2}
      \bigg) H_{0,\{4,1\}}\nonumber\\&&\mbox{}
      +\frac{8 \big(
              1+2 t^2+10 t^4+2 t^6+t^8\big) H_{1,\{4,1\}}}{3 \big(
              1+t^2\big)^4}
      +\bigg(
              -\frac{8 \big(
                      1+2 t^2+10 t^4+2 t^6+t^8\big)}{3 \big(
                      1+t^2\big)^4}
              -\frac{32 t^2 H_1(t)}{3 \big(
                      1+t^2\big)^2}
      \bigg) H_{-1,\{4,1\}}
      \nonumber\\&&\mbox{}
      +\bigg(
              -\frac{4 \big(
                      -3-6 t^2-10 t^4+2 t^6+t^8\big)}{3 \big(
                      1+t^2\big)^4}
              +\frac{32 t^2 H_0}{3 \big(
                      1+t^2\big)^2}
              +\frac{16 t^2 H_1}{3 \big(
                      1+t^2\big)^2}
      \bigg) H_{-1,0}
      +\frac{80 t^2 H_{0,0,1}}{3 \big(
              1+t^2\big)^2}\nonumber\\&&\mbox{}
      -\frac{128 t^2 H_{0,0,\{4,1\}}}{3 \big(
              1+t^2\big)^2}
      +\frac{16 t^2 H_{0,1,1}(t)}{3 \big(
              1+t^2\big)^2}
      -\frac{32 t^2 H_{1,1,\{4,1\}}}{3 \big(
              1+t^2\big)^2}
      -\frac{80 t^2 H_{-1,0,0}}{3 \big(
              1+t^2\big)^2}
      -\frac{16 t^2 H_{-1,0,1}}{3 \big(
              1+t^2\big)^2}
      +\frac{32 t^2 H_{-1,\{4,1\}
      ,1}}{3 \big(
              1+t^2\big)^2}\nonumber\\&&\mbox{}
      +\frac{16 t^2 H_{-1,-1,0}}{3 \big(
              1+t^2\big)^2}
      -\frac{32 t^2 H_{-1,-1,\{4,1\}}}{3 \big(
              1+t^2\big)^2}
      +\frac{4 (-1+t)^2 t^2 (1+t)^2 \log ^2(2)}{3 \big(
              1+t^2\big)^4}
      -\frac{5 t^2 \zeta (3)}{\big(
              1+t^2\big)^2}
      \nonumber\\&&\mbox{}
      +i \pi \Biggl\{
      \frac{\pi ^2 t^2}{3 \big(
              1+t^2\big)^2}
      +\frac{2 t^2 \big(
              2+t^2
      \big)
      \big(1+3 t^2\big)}{3 \big(
              1+t^2\big)^3}
      +\bigg(
              \frac{8 t^4 \big(
                      2+2 t^2+t^4\big)}{3 \big(
                      1+t^2\big)^4}
              +\frac{16 t^2 H_{-1}}{3 \big(
                      1+t^2\big)^2}
      \bigg) H_0\nonumber\\&&\mbox{}
      +\bigg(
              \frac{4 (-1+t) (1+t)}{3 \big(
                      1+t^2\big)}
              +\frac{8 t^2 H_0}{3 \big(
                      1+t^2\big)^2}
      \bigg) H_1
      -\frac{32 t^2 H_0 H_{\{4,1\}}}{3 \big(
              1+t^2\big)^2}
      -\frac{4 (-1+t) (1+t) H_{-1}}{3 \big(
              1+t^2\big)}\nonumber\\&&\mbox{}
      -\frac{8 t^2 H_{0,1}}{\big(
              1+t^2\big)^2}
      +\frac{16 t^2 H_{0,\{4,1\}}}{\big(
              1+t^2\big)^2}
      -\frac{8 t^2 H_{-1,0}}{\big(
              1+t^2\big)^2}
      -\frac{4 t^2 \log (2)}{3 \big(
              1+t^2\big)}
      \Biggr\}
                          \Bigg]
            \,,
\end{eqnarray}
with $t=(1-\sqrt{1-x_c^2})/x_c$ and $x_c=m_c/m_b$.
The expansion of $x_c\to0$ is given by
\begin{eqnarray}
 c_{a, \rm sing}^{(2),s+c}\Big|_{m_2,x_c \to 0} &=&
        n_m C_F T_F \Bigg[
        x_c^2
        \biggl(
                - \frac{\pi^2}{36}
                + \frac{1}{3} \log^2(2)
                - \frac{5}{4}\zeta(3)
                + i \pi        
                \biggl[
                        \frac{\pi^2}{12}
                        - \frac{1}{3} \log(2)
                \biggr]
        \biggr)
        + \frac{\pi^2}{3} x_c^3
        + \mathcal{O}(x_c^4)
        \Bigg]
\end{eqnarray}

Due to the large mass of the top quark it is convenient 
to provide
for $c_{a, \rm sing}^{(2),{b+t}}|_{m_2}$
 only the first few expansion terms in $x_t = m_b/m_t$.  Our
results read
\begin{eqnarray}
        c_{a, \rm sing}^{(2),{b+t}}\Big|_{m_2} &=& 
        n_m C_F T_F \Bigg[
                \frac{55}{24} 
                - \frac{3}{2} \log(x_t)
                + \pi^2
                \biggl(
                        \frac{19}{72}
                        - \frac{2}{3} \log(2)
                \biggr)
                - x_t^2 
                \biggl(
                          \frac{47}{216}
                        + \frac{5}{18} \log(x_t)        
                \biggr)
                \nonumber\\&&\mbox{}
                - x_t^4
                \biggl(
                          \frac{1337}{21600}
                        + \frac{23}{108} \log(x_t)        
                \biggr)
                + \mathcal{O}(x_t^6)
        \Biggr] \,.
\end{eqnarray}

In~\cite{progdata} computer-readable expressions for the four
non-singlet and the three singlet matching coefficients are
provided. We include terms or order $\epsilon$, which are needed for a future
three-loop calculation.

\end{appendix}



\section*{References}

\begin{thebibliography}{99}


%
%

\bibitem{Beneke:2014qea}
M.~Beneke, Y.~Kiyo, P.~Marquard, A.~Penin, J.~Piclum, D.~Seidel and M.~Steinhauser,
Phys. Rev. Lett. \textbf{112} (2014) no.15, 151801
%
[arXiv:1401.3005 [hep-ph]].

\bibitem{Pineda:2011dg}
A.~Pineda,
Prog. Part. Nucl. Phys. \textbf{67} (2012), 735-785
%
[arXiv:1111.0165 [hep-ph]].

\bibitem{Beneke:2013jia}
M.~Beneke, Y.~Kiyo and K.~Schuller,
[arXiv:1312.4791 [hep-ph]].

\bibitem{Beneke:2007gj} 
  M.~Beneke, Y.~Kiyo and K.~Schuller,
  Phys.\ Lett.\ B {\bf 658}, 222 (2008),
  arXiv:0705.4518 [hep-ph].

\bibitem{Czarnecki:1997vz}
A.~Czarnecki and K.~Melnikov,
Phys. Rev. Lett. \textbf{80} (1998), 2531-2534
%
[arXiv:hep-ph/9712222 [hep-ph]].

\bibitem{Beneke:1997jm}
M.~Beneke, A.~Signer and V.~A.~Smirnov,
Phys. Rev. Lett. \textbf{80} (1998), 2535-2538
%
[arXiv:hep-ph/9712302 [hep-ph]].

\bibitem{Marquard:2014pea}
P.~Marquard, J.~H.~Piclum, D.~Seidel and M.~Steinhauser,
Phys. Rev. D \textbf{89} (2014) no.3, 034027
[arXiv:1401.3004 [hep-ph]].

\bibitem{Beneke:2007pj} 
  M.~Beneke, Y.~Kiyo and A.~A.~Penin,
  Phys.\ Lett.\ B {\bf 653}, 53 (2007),
  arXiv:0706.2733 [hep-ph].

\bibitem{Beneke:2014pta}
M.~Beneke, A.~Maier, J.~Piclum and T.~Rauh,
Nucl. Phys. B \textbf{891} (2015), 42-72
%
[arXiv:1411.3132 [hep-ph]].

\bibitem{progdata}
\verb|https://www.ttp.kit.edu/preprints/2021/ttp21-012/|.

\bibitem{Beneke:1997zp} 
  M.~Beneke and V.~A.~Smirnov,
  Nucl.\ Phys.\ B {\bf 522}, 321 (1998)
  [hep-ph/9711391].

\bibitem{Smirnov:2002pj}
V.~A.~Smirnov,
Springer Tracts Mod. Phys. \textbf{177} (2002), 1-262

\bibitem{Smirnov:2019qkx}
A.~V.~Smirnov and F.~S.~Chuharev,
Comput. Phys. Commun. \textbf{247} (2020), 106877
%
[arXiv:1901.07808 [hep-ph]].

\bibitem{Lee:2012cn}
R.~N.~Lee,
arXiv:1212.2685 [hep-ph].

\bibitem{Meyer:2017joq}
C.~Meyer,
Comput. Phys. Commun. \textbf{222} (2018), 295-312
[arXiv:1705.06252 [hep-ph]].

\bibitem{Remiddi:1999ew}
E.~Remiddi and J.~A.~M.~Vermaseren,
Int. J. Mod. Phys. A \textbf{15} (2000), 725-754
[arXiv:hep-ph/9905237 [hep-ph]].

\bibitem{Grozin:2020jvt}
A.~G.~Grozin, P.~Marquard, A.~V.~Smirnov, V.~A.~Smirnov and M.~Steinhauser,
Phys. Rev. D \textbf{102} (2020) no.5, 054008
%
[arXiv:2005.14047 [hep-ph]].

\bibitem{Broadhurst:1991fy}
D.~J.~Broadhurst, N.~Gray and K.~Schilcher,
Z. Phys. C \textbf{52} (1991), 111-122
%

\bibitem{Bekavac:2007tk}
S.~Bekavac, A.~Grozin, D.~Seidel and M.~Steinhauser,
JHEP \textbf{10} (2007), 006
%
[arXiv:0708.1729 [hep-ph]].

\bibitem{Davydychev:1998si}
A.~I.~Davydychev and A.~G.~Grozin,
Phys. Rev. D \textbf{59} (1999), 054023
%
[arXiv:hep-ph/9809589 [hep-ph]].

\bibitem{Fael:2020bgs}
M.~Fael, K.~Sch\"onwald and M.~Steinhauser,
JHEP \textbf{10} (2020), 087
[arXiv:2008.01102 [hep-ph]].

\bibitem{Kniehl:2006qw}
B.~A.~Kniehl, A.~Onishchenko, J.~H.~Piclum and M.~Steinhauser,
Phys. Lett. B \textbf{638} (2006), 209-213
%
[arXiv:hep-ph/0604072 [hep-ph]].

\bibitem{Larin:1993tq}
S.~A.~Larin,
Phys. Lett. B \textbf{303} (1993), 113-118
%
[arXiv:hep-ph/9302240 [hep-ph]].

\bibitem{HarmonicSums}
J.~Vermaseren,
Int. J. Mod. Phys. A \textbf{14} (1999), 2037-2076
%
[arXiv:hep-ph/9806280 [hep-ph]];
E.~Remiddi and J.~Vermaseren,
Int. J. Mod. Phys. A \textbf{15} (2000), 725-754
%
[arXiv:hep-ph/9905237 [hep-ph]];
J.~Bl\"umlein,
Comput. Phys. Commun. \textbf{180} (2009), 2218-2249
%
[arXiv:0901.3106 [hep-ph]];
  J.~Ablinger,
  Diploma Thesis, J. Kepler University Linz, 2009,
  arXiv:1011.1176 [math-ph];
  J.~Ablinger, J.~Bl\"umlein and C.~Schneider,
  J.\ Math.\ Phys.\  {\bf 52} (2011) 102301
  [arXiv:1105.6063 [math-ph]];
J.~Ablinger, J.~Bl\"umlein and C.~Schneider,
J. Math. Phys. \textbf{54} (2013), 082301
%
[arXiv:1302.0378 [math-ph]];
  J.~Ablinger,
  Ph.D. Thesis, J. Kepler University Linz, 2012,
  arXiv:1305.0687 [math-ph];
J.~Ablinger, J.~Bl\"umlein and C.~Schneider,
J. Phys. Conf. Ser. \textbf{523} (2014), 012060
%
[arXiv:1310.5645 [math-ph]];
J.~Ablinger, J.~Bl\"umlein, C.~Raab and C.~Schneider,
J. Math. Phys. \textbf{55} (2014), 112301
%
[arXiv:1407.1822 [hep-th]];
%
J.~Ablinger,
PoS \textbf{LL2014} (2014), 019
%
[arXiv:1407.6180 [cs.SC]];
%
J.~Ablinger,
[arXiv:1606.02845 [cs.SC]];
%
J.~Ablinger,
PoS \textbf{RADCOR2017} (2017), 069
[arXiv:1801.01039 [cs.SC]];
%
J.~Ablinger,
PoS \textbf{LL2018} (2018), 063;
%
%
J.~Ablinger,
[arXiv:1902.11001 [math.CO]].
%

\bibitem{Piclum:2007an}
J.~H.~Piclum,
``Heavy quark threshold dynamics in higher order,''
Dissertation, Hamburg University 2007.

\bibitem{Ablinger:2018zwz}
J.~Ablinger, J.~Bl\"umlein, P.~Marquard, N.~Rana and C.~Schneider,
Nucl. Phys. B \textbf{939} (2019), 253-291
%
[arXiv:1810.12261 [hep-ph]].

\bibitem{Schneider:2007}
C.~Schneider, {S\'em.~Lothar. Combin.\/} {\bf 56} (2007) 1,
 article B56b;
C.~Schneider, in:~{{Computer Algebra in Quantum Field Theory: Integration,
  Summation and Special Functions}\/} Texts and Monographs in Symbolic
  Computation eds. C.~Schneider and J.~Bl\"umlein  (Springer, Wien, 2013) 325
  arXiv:1304.4134 [cs.SC].

\bibitem{ORESYS}
S. Gerhold, {\it Uncoupling Systems of Linear Ore Operator Equations}, Diploma
Thesis, RISC, J. Kepler University, Linz, February 2002.

\bibitem{Eiras:2000rh}
D.~Eiras and J.~Soto,
Phys. Lett. B \textbf{491} (2000), 101-110
[arXiv:hep-ph/0005066 [hep-ph]].

\bibitem{Hoang:2000fm}
A.~H.~Hoang,
[arXiv:hep-ph/0008102 [hep-ph]].

\bibitem{Beneke:2016kkb}
M.~Beneke, Y.~Kiyo, A.~Maier and J.~Piclum,
Comput. Phys. Commun. \textbf{209} (2016), 96-115
%
[arXiv:1605.03010 [hep-ph]].

\bibitem{Ayala:2014yxa}
C.~Ayala, G.~Cveti\v{c} and A.~Pineda,
JHEP \textbf{09} (2014), 045
%
[arXiv:1407.2128 [hep-ph]].

\bibitem{Jegerlehner:2011mw}
F.~Jegerlehner,
Nuovo Cim. C \textbf{034S1} (2011), 31-40
%
[arXiv:1107.4683 [hep-ph]].

\bibitem{Zyla:2020zbs}
P.~A.~Zyla \textit{et al.} [Particle Data Group],
PTEP \textbf{2020} (2020) no.8, 083C01
%

\bibitem{Herren:2017osy}
F.~Herren and M.~Steinhauser,
Comput. Phys. Commun. \textbf{224} (2018), 333-345
doi:10.1016/j.cpc.2017.11.014
[arXiv:1703.03751 [hep-ph]].

\bibitem{Chetyrkin:2009fv} 
  K.~G.~Chetyrkin, J.~H.~Kuhn, A.~Maier, P.~Maierhofer, P.~Marquard, M.~Steinhauser and C.~Sturm,
  Phys.\ Rev.\ D {\bf 80}, 074010 (2009)
  [arXiv:0907.2110 [hep-ph]].

\bibitem{Hatton:2021dvg}
D.~Hatton, C.~T.~H.~Davies, J.~Koponen, G.~P.~Lepage and A.~T.~Lytle,
[arXiv:2101.08103 [hep-lat]].

\bibitem{Beneke:1998rk} 
  M.~Beneke,
  Phys.\ Lett.\ B {\bf 434}, 115 (1998)
  [hep-ph/9804241].

\bibitem{Beneke:2005hg} 
  M.~Beneke, Y.~Kiyo and K.~Schuller,
  Nucl.\ Phys.\ B {\bf 714}, 67 (2005)
  [hep-ph/0501289].

\bibitem{Kniehl:2003ap}
B.~A.~Kniehl, A.~A.~Penin, A.~Pineda, V.~A.~Smirnov and M.~Steinhauser,
Phys. Rev. Lett. \textbf{92} (2004), 242001
[erratum: Phys. Rev. Lett. \textbf{104} (2010), 199901]
%
[arXiv:hep-ph/0312086 [hep-ph]].

\bibitem{Hatton:2020qhk}
D.~Hatton \textit{et al.} [HPQCD],
Phys. Rev. D \textbf{102} (2020) no.5, 054511
%
[arXiv:2005.01845 [hep-lat]].


\end{thebibliography}
\end{document}